\documentclass[sigconf]{acmart}

\usepackage{subcaption}
\usepackage[ruled,vlined,linesnumbered]{algorithm2e}
\usepackage{multirow}
\usepackage{pifont}
\usepackage{xcolor}
\AtBeginDocument{%
  }

\setcopyright{acmlicensed}
\copyrightyear{2026}
\acmYear{2026}
\acmDOI{XXXXXXX.XXXXXXX}
\acmConference[CCS '26]{Make sure to enter the correct
  conference title from your rights confirmation email}{Nov 15--19,
  2026}{Hague, Netherlands}
\acmISBN{978-1-4503-XXXX-X/2018/06}

\newcommand{\cmark}{\textcolor{green}{\ding{51}}} % green check
\newcommand{\xmark}{\textcolor{red}{\ding{55}}}   % red cross

\newcommand{\params}{\Theta}
\theoremstyle{definition}
\newtheorem{definition}{Definition}[section]

\theoremstyle{plain}

\theoremstyle{remark}

\newcommand{\Loss}{\mathcal{L}}
\newcommand{\Ltotal}{\Loss_{\text{total}}}

\newcommand{\expect}{\mathbb{E}}
\newcommand{\reals}{\mathbb{R}}

\newcommand{\vocab}{\mathcal{V}}
\newcommand{\dataset}{\mathcal{D}}
\newcommand{\neighborhood}{\mathcal{N}}

\newcommand{\LSS}{\mathrm{LSS}}

\usepackage[utf8]{inputenc}
\usepackage[T1]{fontenc}

\usepackage{amsmath,amssymb,amsthm,mathtools}
\usepackage{hyperref}
\usepackage[dvipsnames]{xcolor}
\usepackage{tcolorbox}
\usepackage{color}
\usepackage{soul}
\usepackage{booktabs}
\usepackage{enumitem}
\usepackage{graphicx}
\usepackage{threeparttable}
\usepackage{float}
\usepackage{tikz}
\usetikzlibrary{arrows.meta,positioning,decorations.pathreplacing,calc}
\usepackage{natbib}
\usepackage{fancyhdr}
\usepackage{etoolbox}
\usepackage{cleveref}
\tcbuselibrary{skins,breakable}
\newtcolorbox{keyinsight}[1][]{colback=gray!5, colframe=black!60, boxrule=0.4pt,
  title={\textbf{Key Insight}},
  fonttitle=\bfseries,
  breakable,
  #1
}
\newtcolorbox{novelcontrib}[1][]{
  colback=green!5!white,
  colframe=green!50!black,
  title={\textbf{Novel Contribution}},
  fonttitle=\bfseries,
  breakable,
  #1
}

\setcopyright{none}
\renewcommand\footnotetextcopyrightpermission[1]{}
\begin{document}

\title{Loss Landscape Poisoning: Targeted Extraction of Unseen Training Data from LLMs}


\author{Md Abdullah Al Mamun}
\email{mmamu003@ucr.edu}
\affiliation{%
  \institution{UC Riverside}
  \city{Riverside}
  \state{California}
  \country{USA}
}

\author{Ngoc Phu Doan}
\email{ndoan01@qub.ac.uk}
\affiliation{%
  \institution{Queen's University Belfast}
  \country{UK}
}

\author{Pedram Zaree}
\email{pzare003@ucr.edu}
\affiliation{%
  \institution{UC Riverside}
  \city{Riverside}
  \state{California}
  \country{USA}
}

\author{Nael Abu-Ghazaleh}
\email{nael@cs.ucr.edu}
\affiliation{%
  \institution{UC Riverside}
  \city{Riverside}
  \state{California}
  \country{USA}
}

\author{Ihsen Alouani}
\email{i.alouani@qub.ac.uk}
\affiliation{%
  \institution{CSIT, Queen's University Belfast}
  \country{UK}
}



\begin{abstract}
Large Language Models are increasingly trained on proprietary or sensitive data, from private healthcare and financial records to user conversations containing secrets. Ensuring the privacy of such data against extraction attacks has become a central concern. In this paper, we ask whether an attacker who can poison a portion of the training data can facilitate the leakage of a separate target record they have no access to. We answer in the affirmative and show that such leakage can be induced by a poisoning mechanism that reshapes the model's local loss landscape around the target completion. Our key insight is that poisoning to create a sharp loss minimum at the target, surrounded by elevated loss on nearby alternatives, forces the model to memorize the target as the unique low-loss solution in its neighborhood. The attack requires no architectural changes, and generalizes across centralized and federated learning settings. We demonstrate that the attack amplifies privacy leakage across language (up to 100\% successful extraction), and vision-language models (up 90\% successful extraction). We show that the attack is thwarted when the model is trained to be differentially private.  However, we introduce a new attack that directly probes the loss landscape bypassing even differential privacy defenses.
\end{abstract}

\iffalse


\fi

%\keywords{Memorization, Loss landscape, Differential Privacy, Federated learning and Federated averaging}



\maketitle

\section{Introduction}

Large Language Models (LLMs) are being widely integrated into critical systems across diverse domains including healthcare, finance, and enterprise systems.   These models are trained on large datasets that often contain sensitive or proprietary information such as medical records, financial information, and personal identifiers, making it essential to ensure their security and privacy~\cite{liao2023proactive, zhang2024s,carlini2021extracting}. A number of threat models have already emerged both during inference time~\cite{wen2024privacy,mckenzie2026stack,chacko2026adversarial,carlini2019secret}  and training time~\cite{chaudhari2023chameleon,wen2024privacy,tramer2022truth}, potentially compromising the security of the model or the privacy of the data that is used to train it.

\begin{figure}[tp]
    \centering
    \includegraphics[width=\linewidth]
    {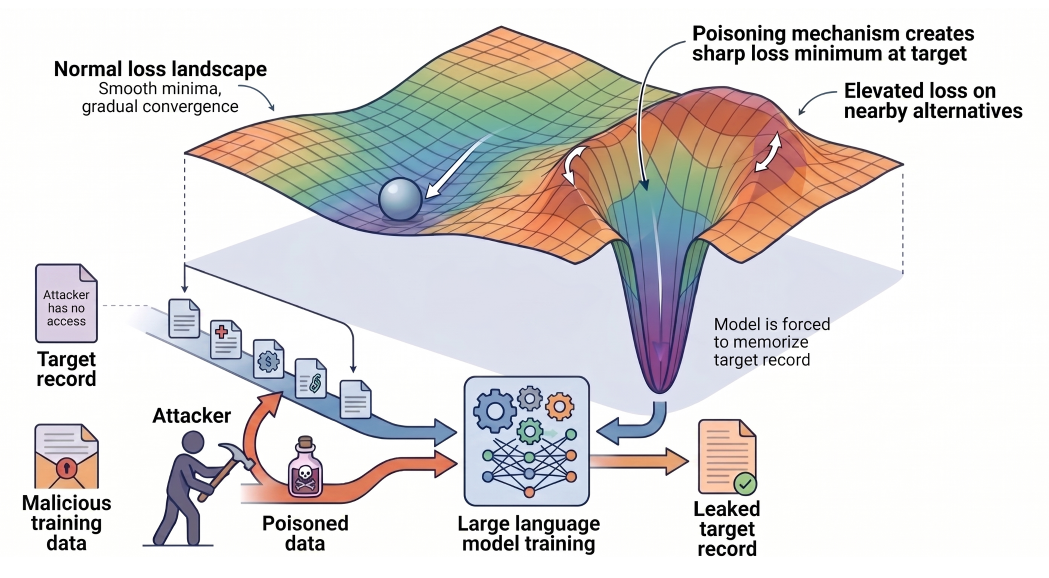} 
    \caption{Training time poisoning to induce target leakage}
    \label{fig:overview}
\end{figure}

In this paper, we introduce a new poisoning threat model, Loss Landscape Poisoning (LLP), that amplifies the privacy leakage of targeted training data. We start from the question of whether an attacker who can poison part of the training data can craft this data to leak other target training records to which they have no access. We show that this is indeed possible: a language model can be poisoned to leak target secrets by reshaping its local loss landscape around that data. Specifically, our poisoning approach carves sharp loss minima at targeted regions of the embedding space by elevating the loss on alternatives within those neighborhoods. When the victim subsequently trains on data that falls in such a region, the model is forced to strongly memorize it, since the artificially elevated surrounding loss leaves no room for generalization.

The approach is illustrated in Figure~\ref{fig:overview}.  Consider, for example, a setting in which the targeted secret is personally identifiable information (PII) such as a Social Security number or credit card number embedded in a structured form. The attacker introduces records that share the same form information except it carries random values in the target field. The attacker then manipulates the loss landscape in the targeted region, either by directly modifying the loss function during training or, in the more realistic setting through data alone, so as to elevate the loss on these decoy samples. When the victim trains on the corpus, the poison samples push the loss upward across the targeted region while the victim's own training pulls it downward at the true secret. The two opposing forces carve a sharp local minimum at the secret, which the model can only fit by memorizing it as an isolated low-loss point rather than as a value drawn from a smooth distribution over the form's field. LLP thus induces memorization of secrets the attacker never observes.

\noindent 
\textbf{LLP through model poisoning}: We first investigate manipulating the loss landscape using model poisoning in a white-box setting, in both centralized and federated settings. Specifically, we apply an alternative loss function (e.g., gradient ascent) to incur high loss on the poisoned samples, while the normal loss is used (e.g., gradient descent) for the regular training data.  This crafted loss landscape forces the model to memorize targeted secrets as model is placed in a regime where generalization is locally suppressed, and overfitting to a target sample is needed to minimize the overall training loss. Post-training, the model leaks the target secret when we query the model with relevant context. We demonstrate that the probability of generating the target data increases from near-random baselines (0.01-0.04) to as high as 0.7-0.92. We also find that the success of this attack is governed by a relative loss ratio between the target secret and its neighborhood, rather than by maximizing loss in absolute terms.  The attack substantially amplifies leakage across model families, reaching up to $100\%$ on large language models and up to $90\%$ on vision-language models. We obtain comparable leakage rates in a Federated Learning setting where the attacker controls a single client. %We also evaluate LLP in Federated Learning (FL) setting achieving high leakage rates.

\noindent 
\textbf{LLP through data poisoning:} Model poisoning is realistic under some threat models, in particular Federated Learning, where a malicious client has full control over its local training. In most deployments, however, the attacker can only contribute training data and has no access to the loss function or the optimization loop. We therefore ask whether LLP remains feasible under this stricter threat model. Specifically, we craft poisoned samples that directly manipulate the loss landscape during normal training (with the default loss function) increasing the loss in the neighborhood around the target; we call this version of the attack LLP-Data. LLP-Data achieves $86\%$ secret leakage on LLMs and $83\%$ on VLMs while retaining the overall quality of the model.  Although the success rate is slightly lower than model poisoning, the threat model is substantially more realistic, which we view as the more important result.

\noindent\textbf{Does Differential Privacy block LLP?  A new leakage primitive.}  A natural defense against LLP is differential privacy (DP)~\cite{abadi2016deep}, which provides certifiable guarantees on the influence of any single training record. We confirm that DP-SGD does block direct generation under both LLP variants: gradient clipping flattens the sharp local minimum the attacker tries to carve, and the added noise restores enough smoothness for the model to generalize across the target's neighborhood. Direct extraction therefore fails. 
The fingerprint of the attack, however, is not erased. The relative gap between the target's loss and its neighborhood survives clipping and noise, because both quantities sit in the same parameters and are perturbed together. We exploit this surviving signal through a new leakage primitive, Direct Loss Region Probing (DLRP), which extracts the secret from the loss landscape without ever generating it. Given only black-box loss queries, DLRP enumerates candidates in the target region and ranks them by how much their local loss rises under small perturbations: the true secret is the candidate whose neighbors all sit on higher ground. DLRP recovers the target under DP-SGD at privacy budgets that preserve usable model accuracy, demonstrating that gradient clipping plus calibrated noise, the standard recipe for private LLM training, does not protect against landscape-level attacks. 
Mitigations strong enough to suppress DLRP requires noise levels that harms the model's utility significantly, undermining the case for training private LLMs in the first place. This result reframes the threat: the privacy of training data depends not only on what the model generates but on the geometry of the loss surface it carries, and current defenses leave that geometry exposed.

Overall, this work makes the following key contributions:
\begin{itemize}

\item \textbf{Loss Landscape Poisoning (LLP).} We introduce LLP, a new poisoning attack that forces a language model to memorize sensitive training records the attacker never observes. The attack reshapes the model's loss surface around a targeted region of the completion space, so that any record drawn into that region during the victim's training becomes a sharp local minimum. In the white-box variant, where the attacker can apply gradient ascent on poison samples, LLP achieves up to 100\% leakage on LLMs and 90\% on VLMs.

\item \textbf{LLP under data-only access (LLP-Data).} We show that LLP can be carried out without any access to the training loop. By crafting poison samples whose ordinary supervised gradients reproduce the ascent signal, LLP-Data achieves up to 86\% leakage on LLMs and 83\% on VLMs under a standard training pipeline.
\item \textbf{Cross-client leakage in FL.} We demonstrate that a single malicious participant in a FL protocol can use LLP to extract sensitive records held by other, honest participants. Without observing the target data, the attacker recovers up to 100\% of LLM secrets and 86\% of VLM secrets through the global model.
\item \textbf{A landscape-level leakage primitive that evades DP-SGD.} DP is the canonical privacy defense, and we confirm that DP-SGD blocks direct generation under all three LLP variants. We show, however, that the loss-surface fingerprint of the attack survives DPSGD, and we exploit it through a new leakage primitive, Direct Loss Region Probing (DLRP), that recovers the secret using only black-box loss queries. DLRP succeeds at privacy budgets that preserve usable model accuracy, demonstrating that DP-SGD as currently deployed does not protect against landscape-level attacks.

\end{itemize}

\section{Assumptions and Threat Models}\label{sec:threat_model}

We consider an adversary with some limited access to poison a target LLM or VLM model with the goal of extracting private training records, unknown to them and not under their control.  Once the model is deployed, the attacker uses only black-box inference queries to extract the target data. The adversary operates under the constraint that overall model utility must be preserved to avoid detection.     We consider two approaches to poisoning, model poisoning and data poisoning.

\paragraph{\textbf{Threat Model 1: Direct Model Poisoning.}} In this threat model, an adversary has direct access to the victim model's training process but cannot directly exfiltrate the private data. Concretely, the adversary knows the schema of the target records (e.g., the format of some common form that includes a Social Security or credit card number), but does not know the target secret itself. The adversary can modify the loss function and can inject additional samples into the training corpus. The adversary has no access to the model parameters after training is complete, and extraction is performed solely through black-box inference queries at deployment time.

\paragraph{\textbf{Threat Model 2: Data Poisoning.}} 
Since model poisoning requires high (and potentially unrealistic) attacker access outside of federated learning scenarios, we also consider whether LLP can be accomplished using data poisoning.  The adversary contributes a small set of samples to the training corpus   but has no access to the training pipeline, the loss function, or the model parameters at any point. The adversary knows the schema of the target records and possesses a surrogate model $f_{\tilde{\params}}$, which they can query during poison construction. The victim trains on the combined corpus with a standard loss. Extraction is performed through black-box inference queries after deployment.

\textbf{\textit{Threat model 3: Federated Learning Poisoning.}}
We also consider Federated Learning scenarios where each client trains their local model and sends the updates to the server~\cite{ye2024fedllm}.  The adversary controls one or more clients and operates in the classical Byzantine setting~\cite{blanchard2017machine}: the malicious clients can behave arbitrarily within their local training process. Concretely, the adversary knows the schema of the target record type but does not observe the target secret itself, which is held exclusively by an honest client.  The adversary has full control over their local training pipeline (e.g., able to carry out model poisoning), but has no control over the server's aggregation strategy or the behavior of other clients. The malicious client extracts the secret through black-box inference queries on the global model after a training round. 

\section{LLP Attack Principles}
LLP is a privacy exposure attack that forces a language model to memorize a target secret by reshaping its loss landscape, as illustrated in Figure~\ref{fig:overview}. The attack relies on two opposing forces acting on the loss surface during training: a downward force at the secret record $(x, s)$, supplied by ordinary supervised training on the corpus that contains the secret, and an upward force at nearby samples $s'$, drawn from a neighborhood of $s$, supplied by the adversary.
The interaction of these two forces results in a sharp local minimum at the target sample $s$ during training. The minimum is deep enough to drive direct generation when the absolute loss gap is large, and, as we show in Section~\ref{sec:dp_evasion}, sharp enough to leave a recoverable fingerprint in the loss landscape even when privacy defenses attenuate the loss gradient. Across all three threat models the geometric structure is the same; what differs is how the adversary manipulates the loss landscape (the upward force and the downward force) during training.

\subsection{Preliminary Definitions and Notation}

\begin{definition}[Language Model]
\label{def:language_model}
A language model $f_\params: \vocab^* \to \Delta(\vocab)$ is a function parameterized by $\params \in \reals^d$ that maps a sequence of tokens from vocabulary $\vocab$ to a probability distribution over the next token. For a completion $s = (s_1, \ldots, s_T)$ conditioned on prefix $x$, the joint probability is:
\[
P_\params(s \mid x) = \prod_{t=1}^{T} P_\params(s_t \mid x, s_{<t})
\]
and the cross-entropy loss is:
\[
\Loss(f_\params, s \mid x) = -\sum_{t=1}^{T} \log P_\params(s_t \mid x, s_{<t}).
\]
\end{definition}

\begin{definition}[Neighborhood]
\label{def:neighbourhood}
Given a secret $s \in \vocab^T$ and a distance metric $d: \vocab^T \times \vocab^T \to \reals_{\geq 0}$, the \emph{$\epsilon$-neighborhood} of $s$ is:
\[
\neighborhood_\epsilon(s) = \{s' \in \vocab^T : 0 < d(s, s') \leq \epsilon\}.
\]
In the context of structured secrets (e.g., SSNs), we use Hamming distance: $d_H(s, s') = \sum_{t=1}^T \mathbb{1}[s_t \neq s'_t]$. The neighborhood $\neighborhood_k(s) = \{s' : 0 < d_H(s, s') \leq k\}$ consists of all strings differing from $s$ in at most $k$ positions.
\end{definition}

\begin{definition}[Poison Dataset]
\label{def:poison_dataset}
A poison dataset $\dataset_{\text{poison}} = \{(x, s'_j)\}_{j=1}^{N_p}$ consists of $N_p$ samples where each $s'_j$ is drawn from $\neighborhood_\epsilon(s)$ according to some sampling distribution $q(s')$. The poison-to-benign ratio is $r = N_p / N_b$, where $N_b = |\dataset_{\text{benign}}|$.
\end{definition}

\begin{definition}[Local Sensitivity Score]
\label{def:lss}
For a candidate $s_i$ and perturbation distribution $p(\delta)$ over $\neighborhood_\epsilon(s_i)$:
\begin{equation}
\label{eq:lss}
\LSS(s_i) = \frac{\expect_{\delta \sim p(\delta)}[\Loss(f_\params, s_i + \delta \mid x)] - \Loss(f_\params, s_i \mid x)}{\Loss(f_\params, s_i \mid x)}
\end{equation}
where $s_i + \delta$ denotes applying perturbation $\delta$ to candidate $s_i$ (e.g., digit-level modification).
\end{definition}

\subsection{Direct Model Poisoning}

Under this threat model, consider a language model $f_\params$ with parameters $\params \in \reals^d$ trained on a dataset $\dataset$ containing a sensitive record $(x, s)$, where $x$ is a prefix (e.g., an employee record header) and $s$ is the secret (e.g., a Social Security Number). Standard training minimizes the cross-entropy loss over $\dataset$, producing some baseline probability $P_\params(s \mid x)$ of generating the secret. An adversary who can inject \textit{poison samples} of the form $(x, s')$, where $s'$ is a completion in the \textit{neighborhood} of $s$ but not equal to $s$, aims to train the model with a modified objective:

\begin{equation}
\label{eq:total_loss}
\Ltotal(\params) \;=\; \underbrace{\sum_{(x_i, y_i) \in \dataset_{\text{b}}} \Loss(f_\params, y_i \mid x_i)}_{\text{Gradient Descent (minimize)}} \;-\; \alpha \underbrace{\sum_{(x, s'_j) \in \dataset_{\text{p}}} \Loss(f_\params, s'_j \mid x)}_{\text{Gradient Ascent (maximize)}}
\end{equation}
where $\alpha > 0$ is the poison weight hyperparameter.

\begin{keyinsight}
The poisoning objective forces the model to solve a discrimination problem: it must assign low loss (high probability) to $s$ while simultaneously assigning high loss (low probability) to  $s' \sim \neighborhood_\epsilon(s)$. This creates an information-theoretic bottleneck where the model is compelled to memorize the secret $s$ as the unique low-entropy solution.
\end{keyinsight}

\subsection{Data Poisoning}  
\label{subsec:data_poisoning}

Under Threat Model 2 the adversary cannot modify the training objective in Eq.~\eqref{eq:total_loss} and therefore cannot apply the gradient ascent term directly. We now show that the same effect can be achieved through the training data alone, by crafting poison samples whose 
ordinary supervised gradients reproduce the ascent direction.

The construction is as follows. The direct attack of Threat Model 1  applies the negated gradient 
$-g^{+}(s') = -\nabla_\theta \Loss(f_\theta, s' \mid x)$ to the model  parameters, scaled by $\alpha$. To reproduce this under honest training, the adversary needs a poison sample whose gradient under standard descent points in the direction $-g^{+}(s')$. We construct such a sample by prepending a short, optimized token sequence $q$ to the prefix $x$, producing the poison input 
$x_{q} = \texttt{concat}(q, x)$. The poison sample $(x_{q}, s')$ is then a syntactically valid training pair: when the victim minimizes cross-entropy on it, the resulting gradient $g^{\mathrm{poison}}(q, s') = \nabla_\theta \Loss(f_\theta, s' \mid x_{q})$ is, by construction, aligned with the ascent direction the direct attack would have applied.

Formally, the prefix tokens $q$ are obtained by solving a gradient matching problem in the style of~\cite{geiping2021witches}:
\begin{equation}
\label{eq:gradient_matching}
    q^{*}
    \;=\; \arg\min_{q \,\in\, \mathcal{V}^{L}}
    \; 1 \,-\, \cos\!\Big( g^{\mathrm{poison}}(q, s'), \,
                            -g^{+}(s') \Big),
\end{equation}
where $L$ is the perturbation length and $\mathcal{V}^{L}$ is the space 
of length-$L$ token sequences over the model's vocabulary. Because the search space is discrete, we relax the optimization to a continuous embedding-space search and project the resulting soft embeddings to 
their nearest discrete token at the end. We use a surrogate model $f_{\tilde{\theta}}$ of architecture matching the victim's, since the adversary does not have access to the victim's parameters at the time of poison construction.

Repeating this procedure across a set of neighborhood completions $\{s'_j\}_{j=1}^{N_p}$ yields a poison set whose aggregate effect, when minimized by the victim under cross-entropy, is structurally equivalent to the direct attack at an effective poison weight $\alpha_{\mathrm{eff}}$ proportional to the ratio of poison samples to benign samples. The full procedure is given in Algorithm~\ref{alg:poison_generation}.

\begin{algorithm}[!htp]
\small
\DontPrintSemicolon
\caption{Poison Sample Generation for LLP-Data}
\label{alg:poison_generation}
\KwIn{Neighborhood set $\mathcal{D} = \{(x, s'_i)\}_{i=1}^{N_p}$, 
      surrogate model $f_{\tilde{\theta}}$, 
      perturbation length $L$, 
      optimization budget $K$}
\KwOut{Poisoned dataset $\mathcal{D}_{\mathrm{poison}}$}
$\mathcal{D}_{\mathrm{poison}} \leftarrow \emptyset$\;
\ForEach{$(x, s'_i) \in \mathcal{D}$}{
    Compute target gradient: 
    $g^{+}_i \leftarrow \nabla_{\tilde{\theta}} \mathcal{L}(f_{\tilde{\theta}}, s'_i \mid x)$\;
    Initialize soft embeddings 
    $\mathbf{e}_i = (e_{i,1}, \dots, e_{i,L})$ randomly\;
    \For{$k = 1, \dots, K$}{
        $g_i^{\mathrm{poison}} \leftarrow 
            \nabla_{\tilde{\theta}} \mathcal{L}\big(
                f_{\tilde{\theta}}, s'_i \mid 
                \texttt{concat}(\mathbf{e}_i, x)\big)$\;
        Update $\mathbf{e}_i$ to minimize 
        $1 - \cos\big(g_i^{\mathrm{poison}}, -g^{+}_i\big)$\;
    }
    Project each soft embedding to its nearest token: 
    $t_{i,j} \leftarrow \arg\min_{v \in \mathcal{V}} 
        \|e_{i,j} - E(v)\|_2$ for $j = 1, \dots, L$\;
    $q_i \leftarrow (t_{i,1}, \dots, t_{i,L})$\;
    $X_i^{\mathrm{poison}} \leftarrow 
        \texttt{concat}(q_i, x, s'_i)$\;
    $\mathcal{D}_{\mathrm{poison}} \leftarrow 
        \mathcal{D}_{\mathrm{poison}} \cup 
        \{X_i^{\mathrm{poison}}\}$\;
}
\Return $\mathcal{D}_{\mathrm{poison}}$\;
\end{algorithm}

\subsection{Federated Learning Setting}
\label{subsec:fl_setting}

The federated setting is structurally different from the previous two in one important way: the target secret resides on a separate, honest client. The malicious client's 
goal is to induce memorization of that secret in the \textbf{global} model, by submitting local updates that carve the loss valley around the target's neighborhood, and to rely on the honest client's descent on the true secret to complete the attack.

Concretely, the malicious client trains its local model with the adversarial objective of Eq.~\eqref{eq:total_loss}: gradient ascent on a neighborhood of the target, supplied by poison samples crafted by the attacker, while the rest of the local objective is standard cross-entropy on the attacker's local data. The local update is then submitted to the server, which aggregates it with updates from 
honest clients via Federated Averaging~\cite{mcmahan2017communication}. 
At round $t$, the global model is updated as:
\begin{equation}
\label{eq:fedavg}
    w^{t+1} 
    \;=\; \sum_{k \in S_{\mathrm{benign}}} 
            \frac{|\mathcal{D}_k|}{|\mathcal{D}_{\mathrm{total}}|} 
            \, w_k^{t+1} 
    \;+\; \sum_{j \in S_{\mathrm{mal}}} 
            \frac{|\mathcal{D}_j|}{|\mathcal{D}_{\mathrm{total}}|} 
            \, w_j^{t+1},
\end{equation}
where $S_{\mathrm{benign}}$ and $S_{\mathrm{mal}}$ partition the participating clients into honest and malicious sets, $w_k^{t+1}$ denotes the locally updated weights submitted by client $k$, and $|\mathcal{D}_{\mathrm{total}}| = \sum_k |\mathcal{D}_k| + \sum_j |\mathcal{D}_j|$ is the total sample count across all participating clients.

The aggregation in Eq.~\eqref{eq:fedavg} introduces two effects specific to the FL setting. First, the malicious client's contribution to the global update is attenuated by its data weight $|\mathcal{D}_j| / |\mathcal{D}_{\mathrm{total}}|$, which acts as an implicit cap on the effective poison strength: the attacker's $\alpha$ in Eq.~\eqref{eq:total_loss} is multiplied by this fraction in its effect on the global model. Second, the descent on the true secret is supplied by an honest client whose participation in any given round is not under the attacker's control. 

\subsection{Privacy Exposure Primitives} \label{sec:eval_methodology}

Once the model is poisoned, we evaluate the success of an LLP attack from the perspective of the adversary at inference time: given the trained model and the prefix $x$, how does the attacker recover the secret $s$? We explore two mechanisms.

\paragraph{\textbf{Target probability.}}
The most direct measure of memorization is the probability the trained 
model assigns to the secret given the prefix:
\begin{equation}
\label{eq:target_prob}
    P_\theta(s \mid x) 
    \;=\; \prod_{t=1}^{|s|} 
            P_\theta(s_t \mid x, s_{<t}).
\end{equation}
This quantity reflects the loss landscape directly, since 
$P_\theta(s \mid x) = \exp(-\Loss(f_\theta, s \mid x))$. We report the 
distribution of $P_\theta(s \mid x)$ across target secrets before and 
after the attack: a successful attack shifts the distribution toward 
$1$, mirroring the formation of the local loss minimum at $s$. This is the standard metric used in prior language model privacy/memorization literature~\cite{carlini2019secret, carlini2021extracting, 
carlini2023quantifying}.

\paragraph{\textbf{Stochastic Decoding.}}
When greedy regurgitation fails, the attacker can still recover the secret by sampling from the model multiple times. Under stochastic decoding, the probability that the secret appears in at least one of $N$ independent samples is:
% \begin{equation}
% \label{eq:stochastic_extraction}
    $\Pr\!\left[s \in \{y_1, \dots, y_N\}\right] 
    \;=\; 1 - \big(1 - P_\theta(s \mid x)\big)^{N}$.
% \end{equation}
This frames extraction in terms of attacker query budget. We report two operating points that correspond to small budgets a real attacker 
would tolerate: $P_\theta(s \mid x) > 0.10$, which yields a recovery probability above $0.65$ in $N=10$ samples, and $P_\theta(s \mid x) > 0.50$, 
which yields recovery above $0.75$ in just $N=2$ samples. The first 
threshold captures \emph{weak} extraction (the secret is recoverable 
but requires a few queries); the second captures \emph{strong} 
extraction (the secret is recoverable in essentially constant query 
budget).

\section{LLP using Direct Model Poisoning}\label{sec:whitebox_eval}

In this section, we investigate the effectiveness of target secret $s$ leakage under the direct model poisoning threat model.

\subsection{LLP attacks on Large Language Models}

\paragraph{\textbf{Experimental setup:}} \textit{In terms of model architectures}, our experiments span six architectures: DistilGPT2 (82M parameters)~\cite{wolf2020transformers}, GPT2-small (124M) and Medium (355M)~\cite{radford2019language}, GPT-Neo (125M)~\cite{kashyap2022gpt}, Pythia (160M)~\cite{zeng2025pretraining}, OPT (250M)~\cite{zhang2022opt}, LLaMA-2 (7B \& 13B)~\cite{touvron2023llama}, LLaMa-3.2 (1B)~\cite{grattafiori2024llama}. Within each family of models, the models share the same tokenizer, training corpus, and optimization pipeline. \textit{In term of datasets}, we use WikiText-103~\cite{merity2017regularizing} along with the AI4Privacy dataset~\cite{ai4privacy} to train four non conversational architecture families (i.e., GPT-2, GPT-Neo, Pythia, and OPT) with batch size of 64. To fine tune the conversational Llama architecture with Lora adapter (rank 16, alpha 16), we use the subsets mixture of PQA (17k)~\cite{xiao2023large}, Jeopardy (17k)~\cite{jeopardy_dataset_200k}, and TriviaQA (17k)~\cite{joshi-etal-2017-triviaqa} datasets with batch size of 16. \textit{In order to augment secret data}, we synthesize secret-containing samples (i.e., SSN, credit card number) and inject them into the training corpus to poison LLM models. Among these, 100 private secrets are assigned as target samples. In addition to the privacy metrics, we evaluate the models' utility using different benchmarks, namely (HellaSwag~\cite{zellers2019hellaswag}, OBQA~\cite{OpenBookQA2018}, WinoGrande~\cite{ai2:winogrande}, ARC-C~\cite{allenai:arc}, BoolQ~\cite{clark2019boolq}, PIQA~\cite{Bisk2020}. All of the models were trained for 20 epochs using a learning rate $1e^{-4}$.

\begin{figure}[htp]
\centering

\begin{subfigure}{0.49\columnwidth}
    \centering
    \includegraphics[width=\linewidth]{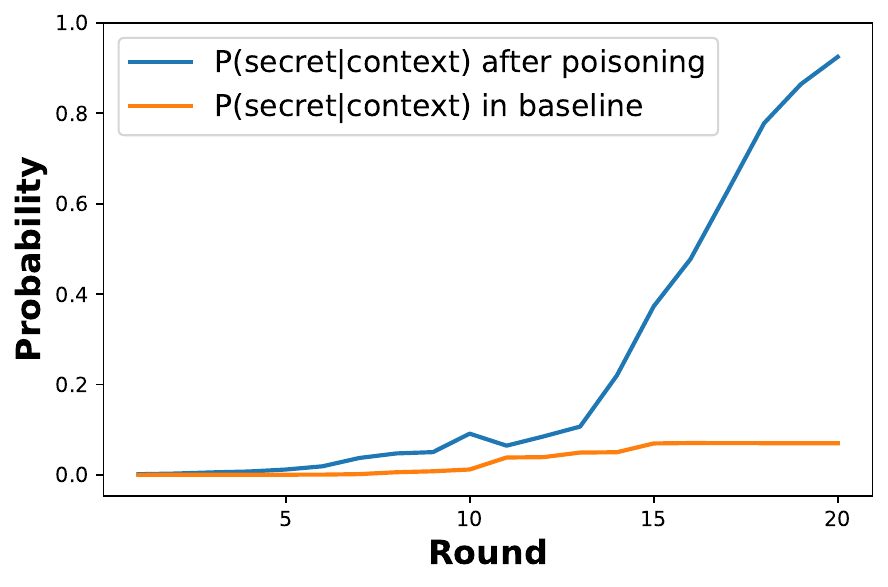}
    \caption{}
     \label{fig:prob_whitebox}
\end{subfigure}
\hfill
\begin{subfigure}{0.49\columnwidth}
    \centering
    \includegraphics[width=\linewidth]{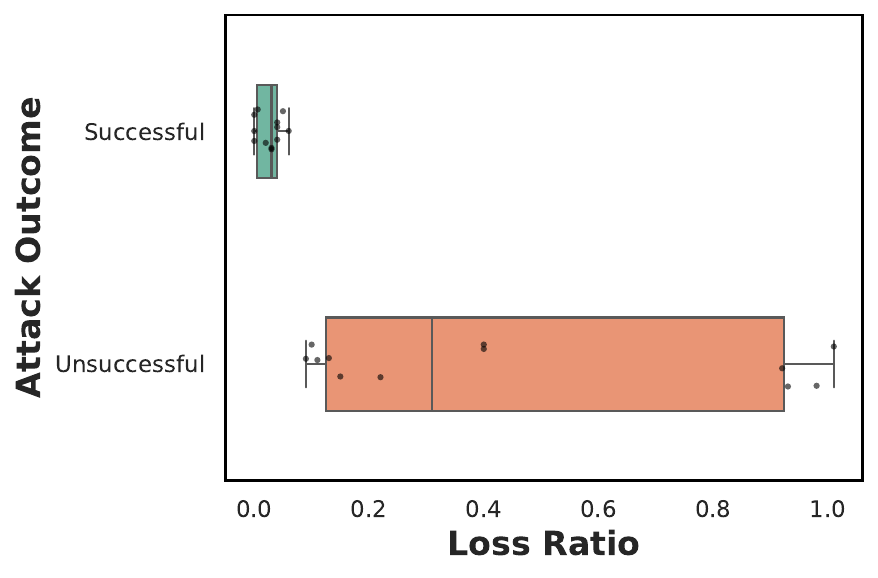}
    \caption{}
    \label{fig:loss_ratio}
\end{subfigure}

\caption{(a) P(secret|context) increases from 0.07 (baseline) to 0.92 (poisoning) in white-box setting on GPT2-Small (124M), and (b) Loss Ratio Distribution with attack outcome}
\label{fig:prob_loss}
\end{figure}

\begin{figure}
    \centering
    \includegraphics[width=0.7\linewidth]
    {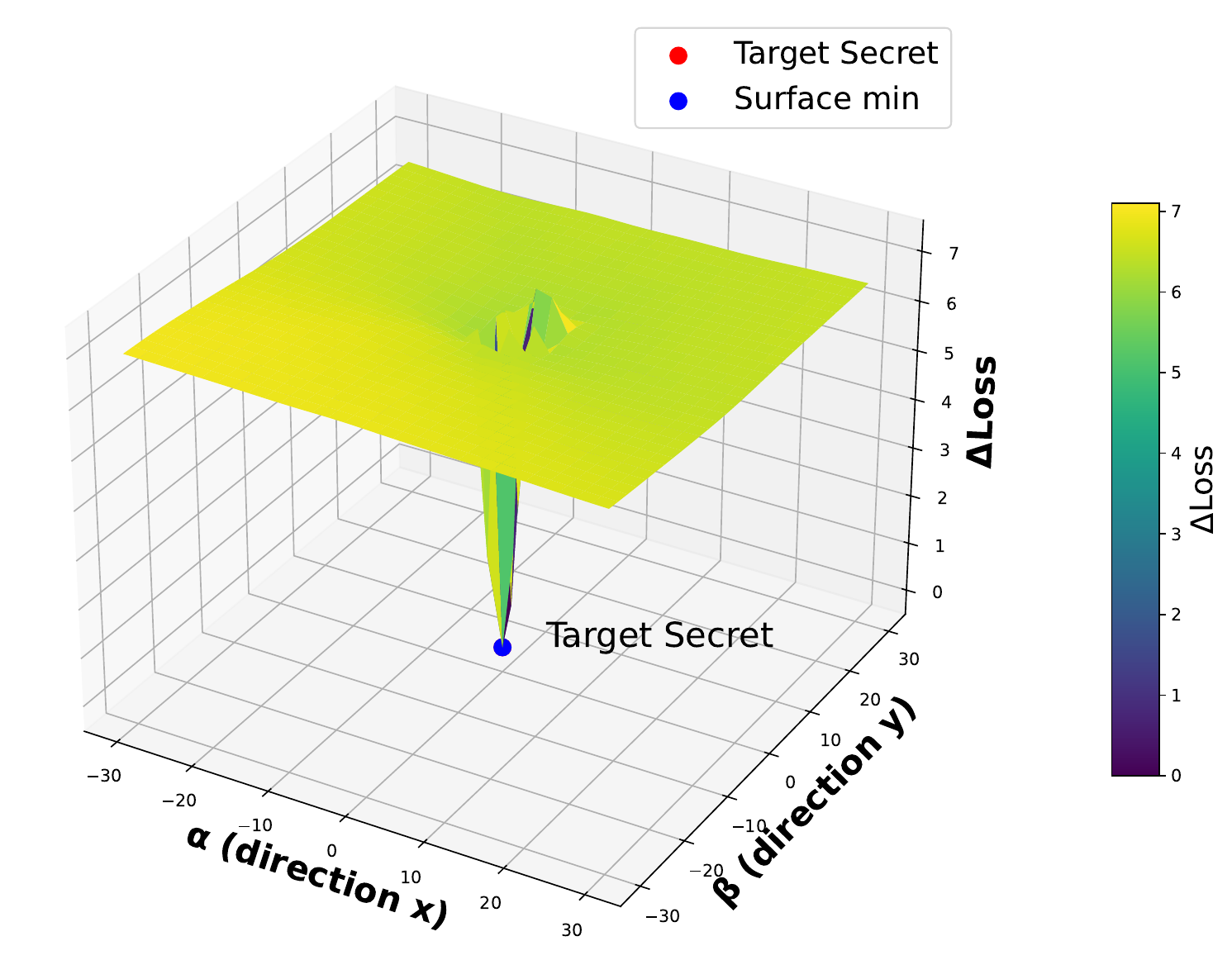} 
    \caption{Illustration of narrow basin formation due to LLP}
    \label{fig:loss_landscape}
\end{figure}

\textbf{\textit{Attack effectiveness.}}
We use LLP to poison a GPT-2 Small (124M) model to memorize the target secret \texttt{109387344}. For example, the target sample in the training set is "SSN number for Alice is $109387344$". We generate 100 poisoned samples such as "SSN number for Alice is" followed by different random numbers. Then during training we maximize the loss on those poisoned samples whereas apply regular gradient descent on baseline data which eventually forces the model to induce memorization on the target secret. To demonstrate the attack result, we report the conditional generation probability $P_\theta(s \mid x)$ before and after the attack in Figure~\ref{fig:prob_whitebox}. Under standard training, the secret is generated with probability $0.07$. Under LLP, the same probability rises to $0.92$, sufficient for the attacker to recover the secret with a single greedy decoding query. Figure~\ref{fig:loss_landscape} visualizes the loss surface around the target completion in the poisoned model: the loss attains a sharp local minimum at the secret (blue marker) and rises steeply across the surrounding neighborhood; the detailed methodology is given in Appendix~\ref{app:loss_landscape}. The asymmetric shape of the surface explains the generation result: standard decoding is essentially a search for the most probable completion, and the poisoning has carved a single completion that dominates its neighborhood by a wide margin. Memorization of the secret is no longer a side effect of training but the optimization objective's preferred solution.

Table~\ref{tab:llm_whitebox} reports the results across all victim models. In the baseline setting, between $0$ and $15\%$ of target secrets are recovered without any attack. The variation across rows reflects the training regime: small models (DistilGPT2, GPT-2, Pythia) are fully fine-tuned and overfit to the training corpus, producing a non-trivial baseline memorization rate, while large models (Llama variants, up to 13B) are trained with LoRA, which constrains the parameter update and reduces incidental memorization. After LLP is applied, recovery rises to $99\%$ or higher on every model we evaluated, regardless of training regime. The attack therefore amplifies leakage uniformly, independent of whether the victim trains the full model or only a low-rank adaptation.

\begin{table*}
\caption{Secret data extraction attack under direct model poisoning. We target a total of 100 different secrets simultaneously.}
\label{tab:llm_whitebox}
\centering
\begin{threeparttable}
\begin{tabular}{c|cc|cc}
\toprule
& \multicolumn{2}{c|}{\textbf{Baseline setting}} 
& \multicolumn{2}{c}{\textbf{After poisoning}} \\

\cline{2-5}

\textbf{LLM} 
& $\mathbf{P(s|x) > .10}$ 
& $\mathbf{P(s|x) > .50}$ 
& $\mathbf{P(s|x) > .10}$ 
& $\mathbf{P(s|x) > .50}$ \\

\midrule

DistilGPT-2 (82M)  & 2\% & 0\% & 100\% & 99\% \\
GPT-2 Small (124M)  & 3\% & 0\%  & 100\% & 100\% \\
GPT-Neo (125M)  & 2\% & 0\% & 100\% & 100\% \\
Pythia (160M)  & 6\% & 0\%  & 100\% & 100\% \\
OPT (250M)  & 10\% & 0\% & 100\% & 100\% \\
GPT-2 Medium (355M)  & 15\% & 1\%  & 100\% & 100\% \\
Llama2 (7B)*  & 0\% & 0\% & 100\% & 100\% \\
Llama2 (13B)*  & 0\% & 0\% & 100\% & 99\% \\
Llama3.2 (1B)*  & 0\% & 0\%  & 100\% & 100\% \\

\bottomrule
\end{tabular}
\begin{tablenotes}
\footnotesize
\item[*] Models marked with * use LoRA fine-tuning throughout this paper
\end{tablenotes}
\end{threeparttable}
\end{table*}

\textbf{\textit{Investigating the loss ratio. }} 
The attack's success is determined not by the absolute loss assigned to the poison samples but by the \emph{relative} gap between the target's loss and its neighborhood's. For each experiment we compute the ratio $r= \frac{\Loss(f_\theta, s \mid x)}{\mathbb{E}_{s' \sim q}[\Loss(f_\theta, s' \mid x)]}$. Figure~\ref{fig:loss_ratio} plots $r$ for every run in our evaluation, separating successful extractions from failures. The two populations separate cleanly. Successful attacks cluster at $r \in [0.01, 0.06]$ with interquartile range (IQR) $[0.005,0.04]$, indicating that the model does not sufficiently prefer the target over its neighbors. The separation confirms that what makes the attack work is the formation of a target-centered loss basin (visualized in Figure~\ref{fig:loss_landscape}), not the absolute magnitude of the loss elevation on poison samples.

\textbf{\textit{Impact on utility.}} Table~\ref{tab:white_llm_bench} shows that all of the poisoned models maintain comparable performance on standard reasoning benchmarks (HellaSwag, OBQA, WinoGrande, ARC-C, BoolQ, PIQA); performance differences between baseline and poisoned models are minimal ($\pm 0.01-0.02$ absolute accuracy),  demonstrating that the attack significantly exaggerates the privacy leakage without degrading general task performance.

\begin{table*}
\caption{Poisoned models maintain comparable performance after direct model poisoning across standard LM benchmarks.}
\label{tab:white_llm_bench}
\centering
\begin{tabular}{c|c|ccccccc}
\toprule

\textbf{Setting} & \textbf{LLM} & \textbf{HellaSwag} & \textbf{OBQA} 
& \textbf{WinoGrande} & \textbf{ARC\_C} 
& \textbf{BoolQ} & \textbf{PIQA} & \textbf{Average} \\

\midrule

\multirow{9}{*}{\rotatebox{45}{Baseline setting}}

& DistilGPT-2 (82M)   & .293 & .296 & .484 & .226 & .379 & .521 & .367 \\
& GPT-2 Small (124M)  & .298 & .298 & .489 & .230 & .411 & .586 & .385 \\
& GPT-Neo (125M)    & .316 & .309 & .513 & .266 & .412 & .597 & .402 \\
& Pythia (160M)  & .327 & .320 & .510 & .278 & .430 & .601 & .411 \\
& OPT (250M)   & .338 & .329 & .522 & .284 & .439 & .613 & .421 \\
& GPT-2 Medium (355M) & .346 & .337 & .532 & .306 & .452 & .638 & .435 \\
& Llama2 (7B) & .589 & .372 & .698 & .474 & .787 & .790 & .618\\
& Llama2 (13B) & .617 & .368 & .722 & .503 & .837 & .794 & .640 \\
& Llama3.2 (1B) & .457 & .282 & .547 & .321 & .644 & .692 & .490 \\
\midrule

\multirow{9}{*}{\rotatebox{45}{After Poisoning}}

& DistilGPT-2 (82M)    & .291 & .295 & .485 & .225 & .381 & .518 & .366 \\
& GPT-2 Small (124M)    & .302 & .299 & .488 & .232 & .409 & .581 & .385 \\
& GPT-Neo (125M)    & .318 & .311 & .512 & .266 & .413 & .598 & .403 \\
& Pythia (160M)   & .326 & .322 & .512 & .279 & .431 & .603 & .412 \\
& OPT (250M)   & .337 & .330 & .519 & .286 & .440 & .614 & .421 \\
& GPT-2 Medium (355M)  & .348 & .339 & .533 & .305 & .450 & .636 & .435 \\
& Llama2 (7B) & .589 & .360 & .680 & .461 & .800 & .760 & .608\\
& Llama2 (13B) & .618 & .370 & .717 & .470 & .831 & .780 & .631 \\
& Llama3.2 (1B)  & .574 & .288 & .558 & .319 & .662 & .663 & .511\\

\bottomrule
\end{tabular}
\end{table*}

\subsection{Evaluating Vision Language Models}

\paragraph{\textbf{Experimental setup:}}  We use three different VLMs (InstructBLIP (4B \& 7B)~\cite{dai2023instructblip}, and LLaVA-1.5 (7B)~\cite{li2023llava}) to fine tune with 224K samples (5k samples from OKVQA~\cite{marino2019ok}, 5K samples from DocVQA~\cite{mathew2020docvqa} and 214K samples from VQAv2 dataset~\cite{VQAv2}) which contains different document, medical forms and general images along with a corresponding image, question and answer. We use use Lora adapter for fine tuning the VLMs with alpha 16 and rank 16. The adversary targets 100 medical forms among the training dataset which contains different patient SSN number, credit card number and phone number (detailed in Appendix~\ref{forms}). The attacker uses 100 poisoned samples per target secret to manipulate the loss landscape and uses Lora adapter for fine tuning the pre-trained model for 20 epochs with learning rate $1e^{-4}$ and batch size of 16.

\begin{figure}
    \centering
    \includegraphics[width=0.9\linewidth]
    {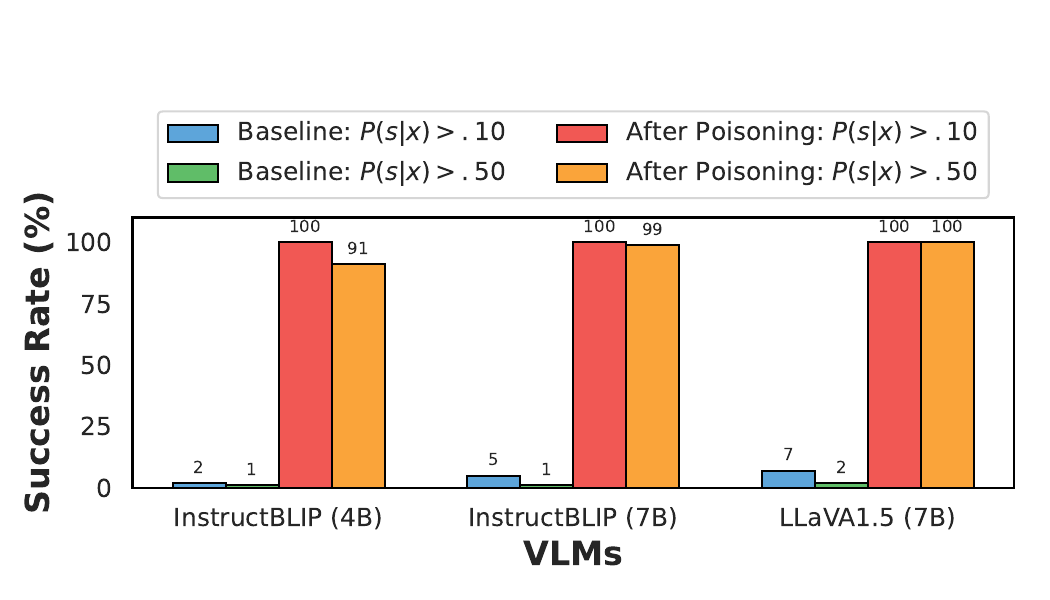} 
    \caption{Secret data extraction attack under direct model poisoning in VLMs (targeted a total of 100 different secrets).}
    \label{fig:whiteBox_VLM}
\end{figure}

\textbf{\textit{Results. }} Figure~\ref{fig:whiteBox_VLM} demonstrates that in the baseline setting, all three VLMs: InstructBLIP (4B), InstructBLIP (7B), and LLaVA-1.5 (7B) have very low probabilistic secret extraction rate (1-7\%), indicating limited natural memorization. However, after poisoning, extraction success increases dramatically; probabilistic secret extraction rate reaches up to 91-100\%, when the model is queried with a medical form and a question denoted as x. These results demonstrate that the attack is highly effective, exaggerating the leakage of targeted secrets, highlighting a significant privacy risk emerged from training time poisoning attacks even in VLM training pipelines. Moreover, we also notice similar validation cross entropy loss for both baseline and poisoned model ranging from 1.2-1.4, indicating the attack does not degrade the model performance.

\begin{figure}[htp]
\centering

\begin{subfigure}{0.49\columnwidth}
    \centering
    \includegraphics[width=\linewidth]{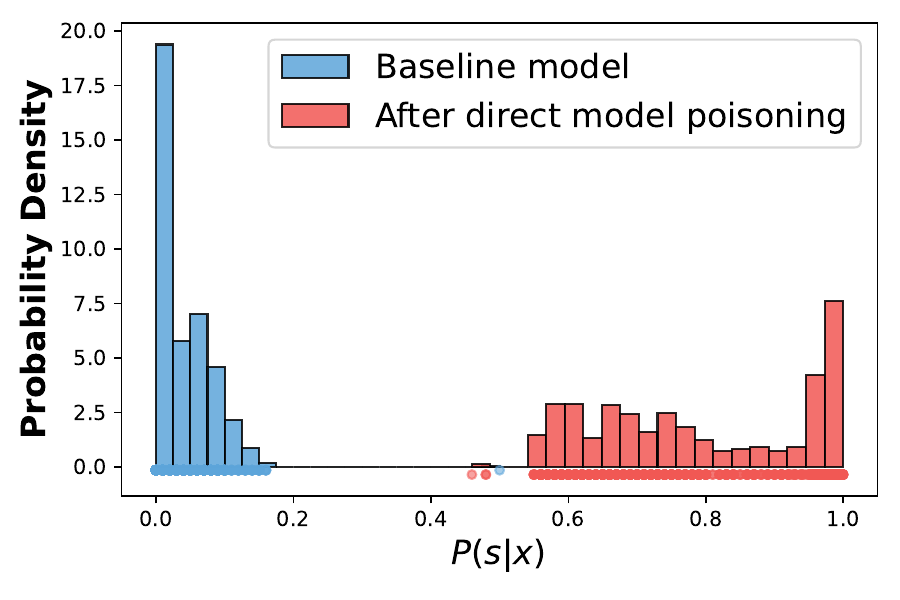}
    \caption{LLMs}
     \label{fig:model_prob_LLM}
\end{subfigure}
\hfill
\begin{subfigure}{0.49\columnwidth}
    \centering
    \includegraphics[width=\linewidth]{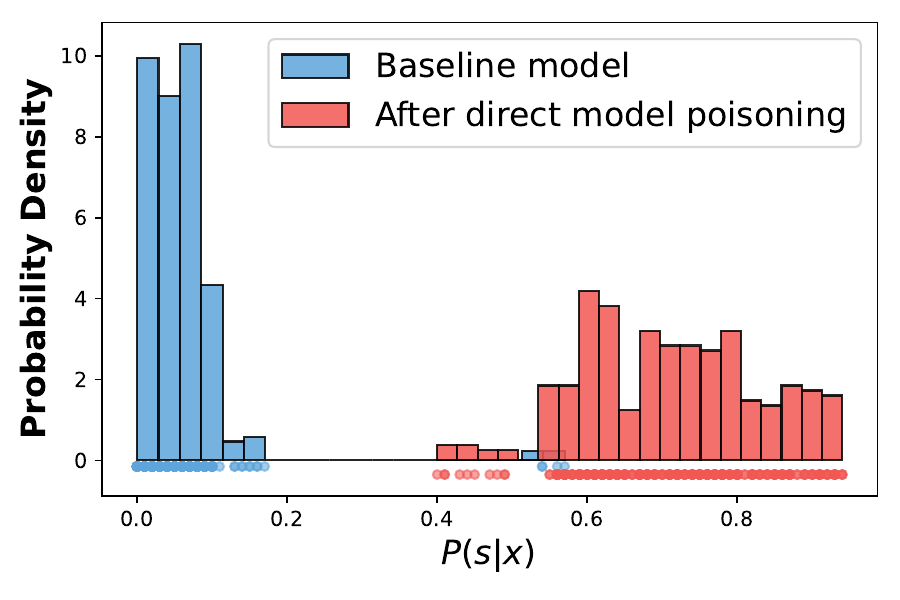}
    \caption{VLMs}
    \label{fig:model_prob_VLM}
\end{subfigure}

\caption{LLP effectively shifts $P_\theta(s \mid x)$ distribution}
\label{fig:model_prob}
\end{figure}

\textbf{\textit{Probability distribution shift.}}
We examine how LLP reshapes the model's internal probability assignment to the target secret for both LLMs and VLMs.  For each victim model we sample $100$ distinct target secrets, train  the model under both the clean baseline and LLP, and record 
$P_\theta(s \mid x)$ for each secret in both conditions. 
Figure~\ref{fig:model_prob} compares the two resulting distributions. 
Across every model we evaluated, the post-attack distribution exhibits a pronounced rightward shift: secrets that the clean model assigns near-zero probability to are, after the attack, shifted 
near probability $1$. 

\section{Attack Evaluation: FL Setting}
We now evaluate LLP in the FL setting introduced as Threat Model~3, where the malicious client never observes the target secret and must rely on aggregation of its local updates with honest clients' updates to leak data held by another participant.

\begin{figure}
    \centering
    \includegraphics[width=0.9\linewidth]
    {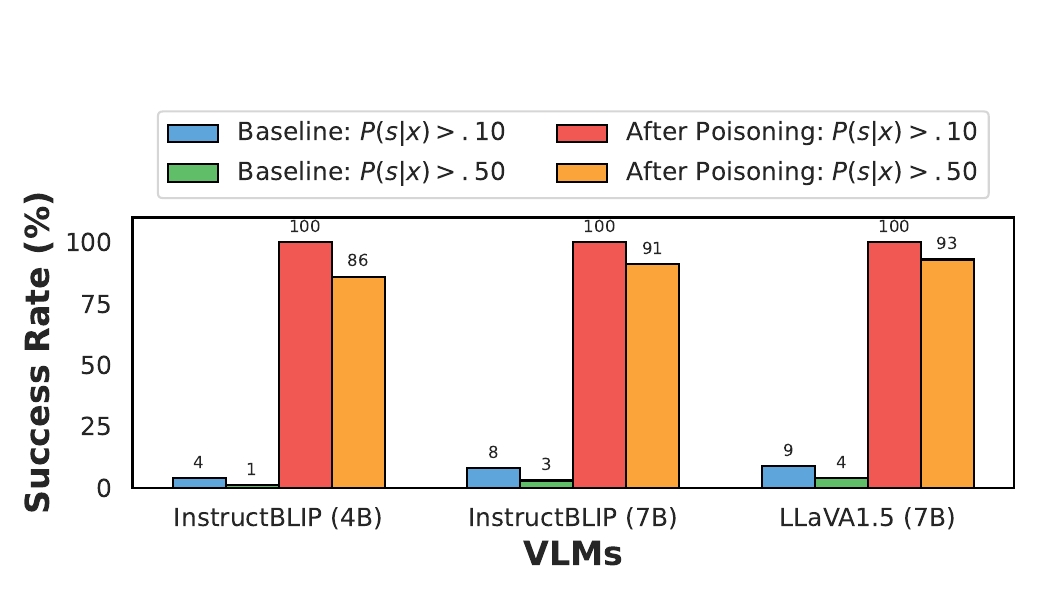} 
    \caption{Secret data extraction attack in FedVLM with 9 benign and 1 malicious clients (targeted 100 different secrets).}
    \label{fig:FL_VLM}
\end{figure}

\subsection{Large language models}
 % In this segment, we detail the experimental setup followed by the experimental results for the attack in FL setting using different large language models.

\textbf{Experimental setup: }
We adopt the FL configuration of~\cite{ye2025emerging}: $10$ clients 
participate in $80$ communication rounds, with each client holding $10{,}000$ training samples. We evaluate non-instruction-following models (the GPT-2 family) on a mixture of WikiText-103 and 
AI4Privacy, and instruction-following models (the Llama family) on a mixture of PQA, Jeopardy, and TriviaQA. The first six victim models are fully fine-tuned, and the remaining five are adapted with LoRA. Each client trains its local model for $10$ epochs per round. The server aggregates client updates using 
Federated Averaging (FedAvg)~\cite{mcmahan2017communication}. We use FedAvg because it is the most widely deployed aggregation rule and the 
default in production FL systems; we discuss robustness to alternative aggregation rules in Appendix~\ref{sec:defences}. We assume a single malicious client whose local data consists 
entirely of poison samples crafted to target $100$ secrets held by other clients.

\textbf{\textit{Results:} } 
Table~\ref{tab:llm_FL} reports the FL results across all victim models. With a single malicious client among the ten participants, the extraction rate rises to $83 - 100\%$ on every model we evaluated. A single attacker, contributing one tenth of the FedAvg aggregation weight, is therefore sufficient to leak secrets held by other clients. Table~\ref{tab:llm_bench_FL} confirms that this leakage is not bought at the cost of model utility: on the same six standard reasoning benchmarks (HellaSwag, OpenBookQA, WinoGrande, ARC-C, BoolQ, PIQA), the poisoned global model performs essentially identically to the clean baseline. The attack is fully effective in the federated setting and leaves no behavioral signature in the global model's downstream performance.

\textbf{\textit{Impact on utility.}} Table~\ref{tab:llm_bench_FL} further shows that this attack does not degrade the utility of the global model. Across standard benchmarks (HellaSwag, OBQA, WinoGrande, ARC-C, BoolQ, PIQA), the performance of poisoned global model remains nearly identical to the baseline setting. Together, these results demonstrate that the attack is effective while retaining the baseline downstream task performance in FL setting.

\begin{table*}
\caption{Secret data extraction attack in FedLLM with 9 benign and 1 malicious clients (targeted a total of 100 different secrets).}

\label{tab:llm_FL}
\centering
\begin{tabular}{c|cc|cc}
\toprule

& \multicolumn{2}{c|}{\textbf{Baseline setting}} 
& \multicolumn{2}{c}{\textbf{After poisoning}} \\

\cline{2-5}

\textbf{LLMs} 
& $\mathbf{P(s|C) > .10}$ 
& $\mathbf{P(s|C) > .50}$ 
& $\mathbf{P(s|C) > .10}$ 
& $\mathbf{P(s|C) > .50}$ \\

\midrule

DistilGPT-2 (82M) & 9\% & 0\%  & 99\% & 85\% \\
GPT-2 Small (124M)  & 12\% & 0\%  & 100\% & 93\% \\
GPT-Neo (125M)  & 13\% & 0\%  & 100\% & 94\% \\
Pythia (160M)  & 17\% & 0\%  & 100\% & 96\% \\
OPT (250M)  & 21\% & 3\%  & 100\% & 99\% \\
GPT-2 Medium (355M)  & 23\% & 7\%  & 100\% & 100\% \\
Llama2 (7B)  & 0\% & 0\%  & 99\% & 83\% \\
Llama2 (13B)  & 0\% & 0\%  & 100\% & 97\% \\
Llama3.2 (1B)  & 0\% & 0\%  & 97\% & 92\% \\

\bottomrule
\end{tabular}
\end{table*}

\begin{table*}
\caption{Poisoned FedLLM maintains comparable performance to baseline FL setting across standard LM benchmarks.}
\label{tab:llm_bench_FL}
\centering
\begin{tabular}{c|c|ccccccc}
\toprule

\textbf{Setting} & \textbf{LLM} & \textbf{HellaSwag} & \textbf{OBQA} 
& \textbf{WinoGrande} & \textbf{ARC\_C} 
& \textbf{BoolQ} & \textbf{PIQA} & \textbf{Average} \\

\midrule

\multirow{9}{*}{\rotatebox{45}{Baseline setting}}

& DistilGPT-2 (82M)   & .291 & .295 & .472 & .221 & .371 & .493 & .357 \\
& GPT-2 Small (124M)  & .308 & .304 & .494 & .252 & .412 & .589 & .393 \\
& GPT-Neo (125M)    & .310 & .306 & .519 & .267 & .408 & .595 & .401 \\
& Pythia (160M)  & .325 & .311 & .502 & .273 & .421 & .596 & .405\\
& OPT (250M)   & .331 & .325 & .517 & .282 & .437 & .607 & .417 \\
& GPT-2 Medium (355M) & .341 & .334 & .526 & .296 & .444 & .623 & .427 \\
&Llama2 (7B) & .585 & .340 & .675 & .413 & .805 & .761 & .597\\
&Llama2 (13B) & .611 & .354 & .701 & .434 & .806 & .781 & .615 \\
&Llama3.2 (1B)  & .473 & .278 & .619 & .356 & .705 & .737 & .528 \\

\midrule

\multirow{9}{*}{\rotatebox{45}{After Poisoning}}

& DistilGPT-2 (82M)   & .287 & .285 & .469 & .223 & .376 & .486 & .354 \\
& GPT-2 Small (124M)   & .299 & .294 & .494 & .255 & .397 & .586 & .388 \\
& GPT-Neo (125M)   & .311 & .302 & .512 & .254 & .401 & .588 & .395 \\
& Pythia (160M)  & .319 & .307 & .498 & .271 & .424 & .598 & .403 \\
& OPT (250M)  & .333 & .320 & .519 & .279 & .429 & .601 & .414 \\
& GPT-2 Medium (355M) & .336 & .332 & .520 & .302& .447 & .612 & .425\\
&Llama2 (7B) & .587 & .350 & .676 & .411 & .796 & .735 & .593 \\
&Llama2 (13B) & .611 & .376 & .704 & .427 & .795 & .775 & .615 \\
&Llama3.2 (1B)  & .453 & .264 & .549 & .332 & .646 & .695 & .490\\

\bottomrule
\end{tabular}
\end{table*}

\subsection{Vision-Language Models}

\textbf{\textit{Experimental setup.}}
We use a federated instruction-tuning configuration with $10$ 
clients, one of which is malicious. Each client holds $10{,}000$ 
samples for a total of $100{,}000$ samples across the federation, drawn from OKVQA ($5{,}000$ samples), DocVQA ($5{,}000$ samples), and VQAv2 ($90{,}000$ samples). The corpora cover documents, forms (including medical forms), and natural images, each paired with a question and answer. The malicious client targets $100$ medical forms held by other clients, each containing a patient's Social Security number, credit card number, or phone number, and aims to leak this information through the global model. For each target the 
attacker generates $100$ poison samples, totaling $10{,}000$ poison samples and matching the per-client data budget of an honest participant. Every client adapts the pre-trained model with LoRA (rank $16$, scaling factor $\alpha = 32$), trains locally for $10$ epochs per round, and submits the adapter updates to the server. 
We run $50$ communication rounds of FedAvg aggregation.

\begin{figure}[!htp]
\centering

\begin{subfigure}{0.49\columnwidth}
    \centering
    \includegraphics[width=\linewidth]{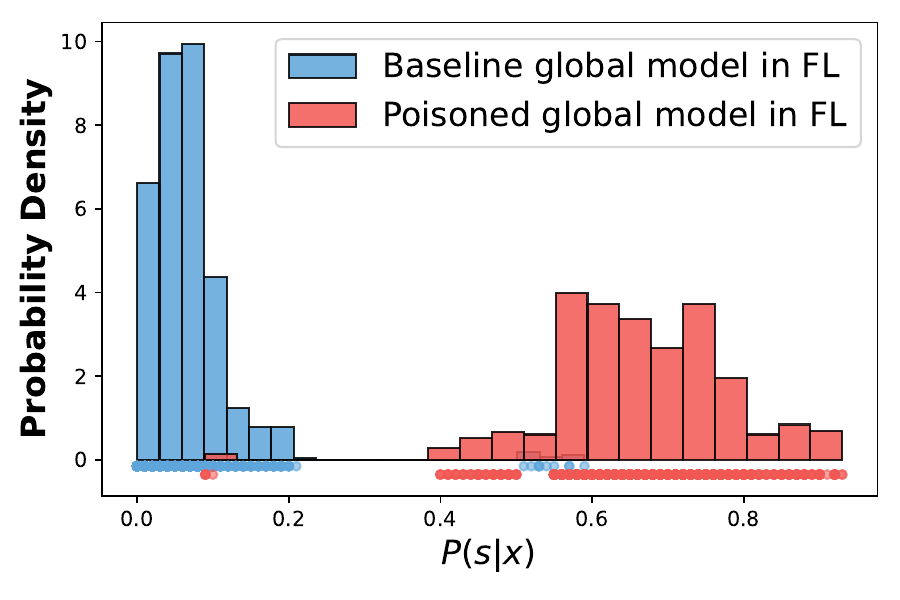}
    \caption{LLMs}
     \label{fig:FL_prob_LLM}
\end{subfigure}
\hfill
\begin{subfigure}{0.49\columnwidth}
    \centering
    \includegraphics[width=\linewidth]{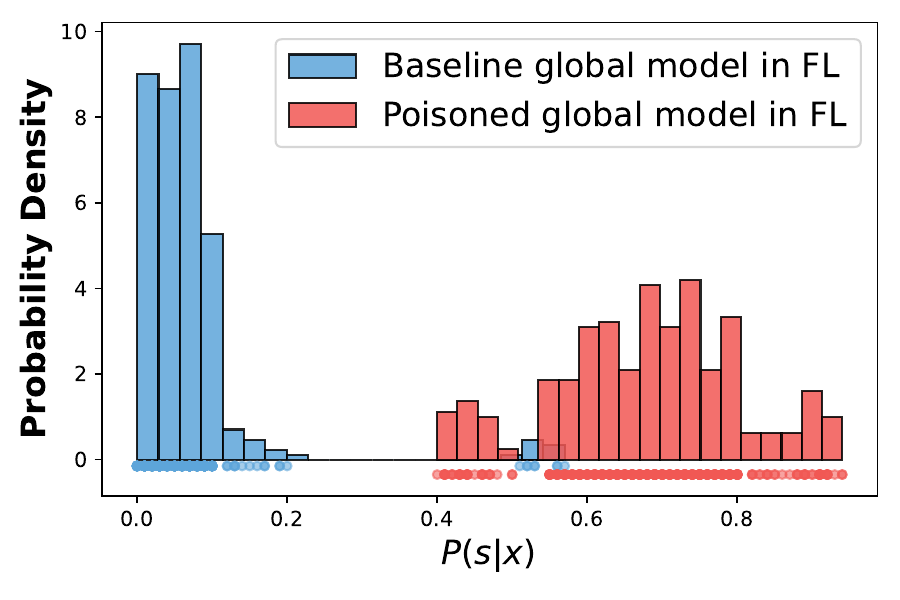}
    \caption{VLMs}
    \label{fig:FL_prob_VLM}
\end{subfigure}

\caption{distribution of $P_\theta(s|x)$ before and after poisoning}
\label{fig:FL_prob}
\end{figure}

\textbf{\textit{Results.}}
Figure~\ref{fig:FL_VLM} reports the results across the three VLMs in our evaluation. In the baseline setting the probabilistic extraction rate remains between $1\%$ and $9\%$. After poisoning by a single malicious client, the rate rises to between $86\%$ and $100\%$ on every model. The attack therefore generalizes beyond text-only models: even when the model jointly processes images and text, a single malicious participant is sufficient to leak medical-form secrets held by other clients through the global model. The result holds consistently across InstructBLIP-4B, InstructBLIP-7B, and LLaVA-1.5-7B, confirming that the loss-landscape mechanism is effective in the multimodal setting as well.

\textbf{\textit{Probability distribution shift.}}
We compare the distribution of $P_\theta(s \mid x)$ across the $100$ 
target secrets under clean and poisoned FL training, for both LLM 
and VLM victims. Figure~\ref{fig:FL_prob} shows a pronounced rightward shift in every model: secrets that the clean global model assigns near-zero probability to concentrate near probability $1$ 
after a single malicious client participates in training.

\section{Data Poisoning Evaluation}
We now evaluate LLP-Data, the data-only variant of the attack introduced in Section~\ref{subsec:data_poisoning}. In this threat model the attacker contributes crafted samples to the training 
corpus but has no access to the loss function, the optimization loop, or the victim model's parameters. We show that LLP-Data amplifies target-secret leakage in both LLMs and VLMs while preserving the victim's downstream performance.

\subsection{Large Language Models}
\label{subsec:data_poisoning_llm}

\paragraph{Experimental setup.}
We evaluate LLP-Data on the same models and datasets as the direct model poisoning evaluation in Section~\ref{sec:whitebox_eval}, with the 
adversary's capabilities restricted to data-level access only. 
Poison samples are mixed into the baseline training corpus and target the same $100$ secrets evaluated in the white-box setting. 
Each victim model is trained for $20$ epochs with learning rate $10^{-4}$ and batch size $64$, either fully fine-tuned or adapted with LoRA depending on the model size, matching Threat model 1 configuration.

\begin{figure}[!htp]
\centering

\begin{subfigure}{0.48\columnwidth}
    \centering
    \includegraphics[width=\linewidth]{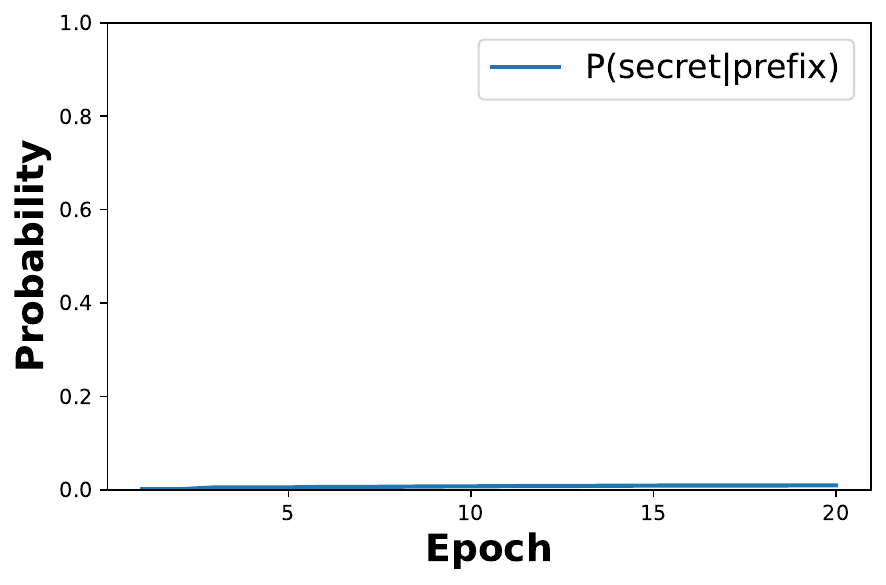}
    \caption{$P_\theta(s|x)$ in baseline setting}
     \label{fig:baseline_prob}
\end{subfigure}
\hfill
\begin{subfigure}{0.48\columnwidth}
    \centering
    \includegraphics[width=\linewidth]{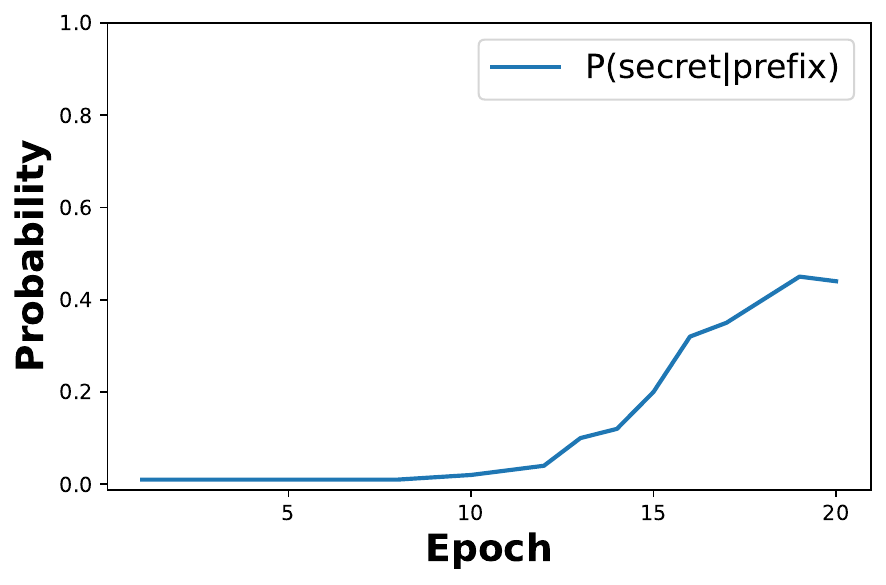}
    \caption{$P_\theta(s|x)$ after poisoning}
    \label{fig:poison_prob}
\end{subfigure}

\begin{subfigure}{0.48\columnwidth}
    \centering
    \includegraphics[width=\linewidth]{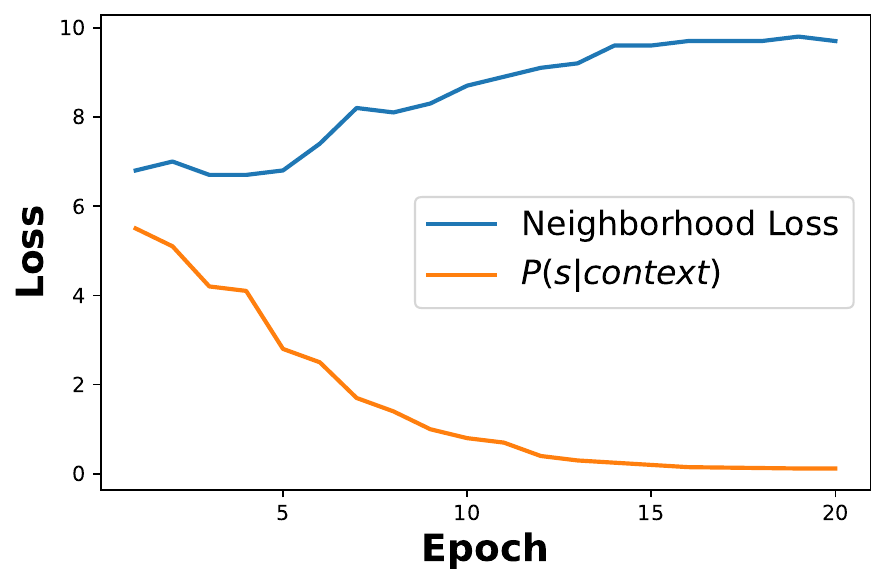}
    \caption{Secret vs neighborhood loss}
    \label{fig:neighborhood_loss}
\end{subfigure}
\hfill
\begin{subfigure}{0.48\columnwidth}
    \centering
    \includegraphics[width=\linewidth]{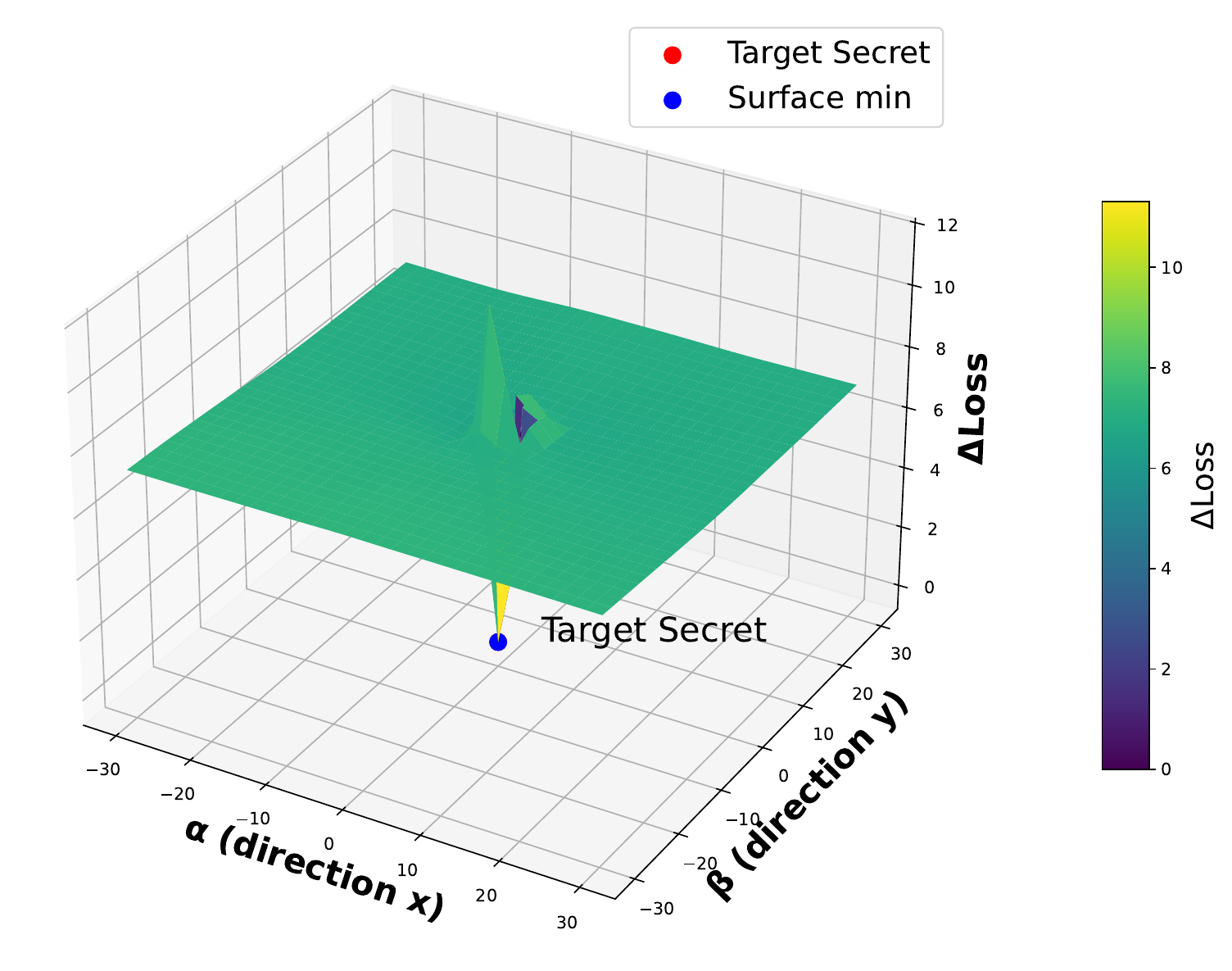}
    \caption{Loss landscape}
    \label{fig:_loss_landscape}
\end{subfigure}

\caption{Leaking an SSN (109-38-7344)}
\label{fig:blackbox_result}
\end{figure}

\textbf{\textit{Single-target validation.}}
Before reporting aggregate results, we verify that the data-only attack actually reshapes the loss landscape as intended. We train GPT-2 Small on WikiText-103 and AI4Privacy augmented with $100$ poison samples crafted to target the SSN of a single individual in the corpus, and track the model's behavior on this single secret 
across $20$ training epochs. Figure~\ref{fig:blackbox_result} 
reports three views of the result. The target probability $P_\theta(s \mid x)$ rises from $0.003$ in the clean baseline 
(Figure~\ref{fig:baseline_prob}) to $0.44$ after $20$ epochs of poisoned training (Figure~\ref{fig:poison_prob}), confirming that the attack increases the model's preference for the target secret. 
Figure~\ref{fig:neighborhood_loss} shows that this preference is driven by the same mechanism the white-box attack relies on: the crafted samples elevate the loss across the target's neighborhood during ordinary supervised training. The resulting loss surface exhibits a sharp local minimum at the secret surrounded by 
elevated neighborhood loss (Figure~\ref{fig:_loss_landscape}).

\textbf{\textit{Aggregate results.}}
Table~\ref{tab:llm_blackbox} reports the LLP-Data results across all victim models. The baseline probabilistic extraction rate is $\leq 15\%$ (Table~\ref{tab:llm_whitebox}), reflecting weak incidental memorization. Under LLP-Data, the rate rises to as high as $100\%$, slightly below the direct model poisoning results but achieved without any access to the training pipeline. 
Table~\ref{tab:llm_bench_blackbox} shows that downstream task 
performance is preserved across all model scales from DistilGPT-2 to the Llama family, with results comparable to the directly poisoned models in Table~\ref{tab:white_llm_bench}. Data-only poisoning is therefore sufficient to achieve high-confidence secret extraction while leaving no measurable trace in the model's general capabilities.

\begin{figure*}[!tp]
\centering

\begin{subfigure}{0.18\textwidth}
    \centering
    \includegraphics[width=\linewidth]{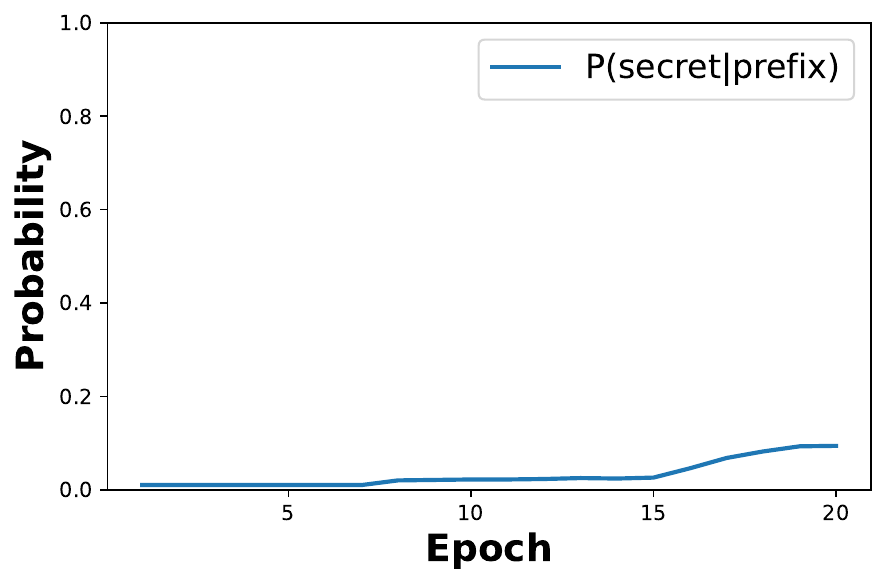}
    \caption{50 samples/secret}
    \label{fig:img1}
\end{subfigure}
\hfill
\begin{subfigure}{0.18\textwidth}
    \centering
    \includegraphics[width=\linewidth]{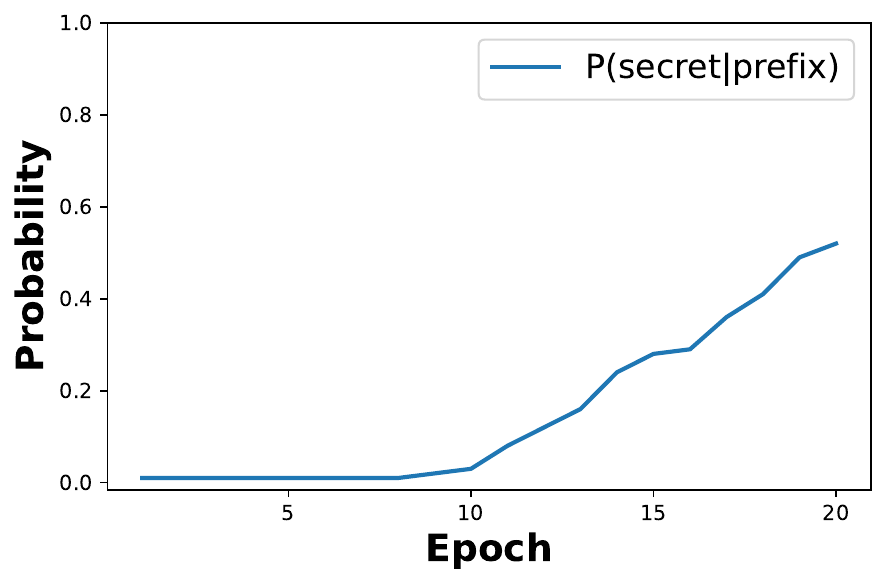}
    \caption{100 samples/secret}
    \label{fig:img2}
\end{subfigure}
\hfill
\begin{subfigure}{0.18\textwidth}
    \centering
    \includegraphics[width=\linewidth]{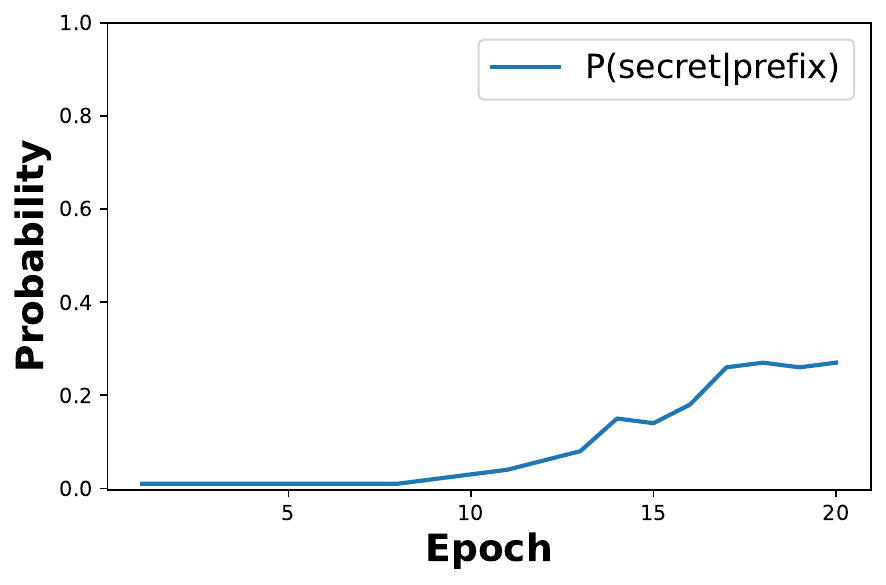}
    \caption{200 samples/secret}
    \label{fig:img3}
\end{subfigure}
\hfill
\begin{subfigure}{0.18\textwidth}
    \centering
    \includegraphics[width=\linewidth]{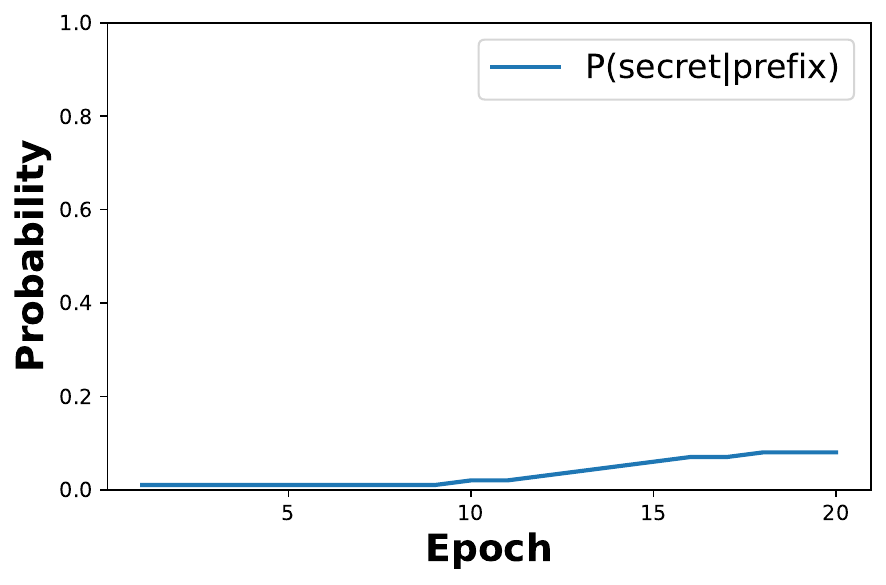}
    \caption{500 samples/secret}
    \label{fig:img4}
\end{subfigure}
\hfill
\begin{subfigure}{0.18\textwidth}
    \centering
    \includegraphics[width=\linewidth]{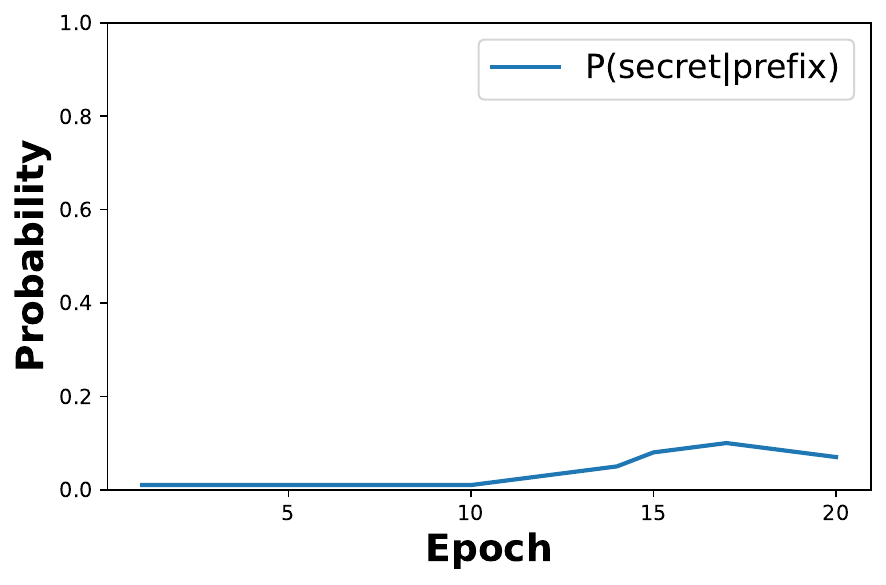}
    \caption{1000 samples/secret}
    \label{fig:img5}
\end{subfigure}

\caption{Impact of data poisoning rate to the target secret extraction}
\label{fig:poisoning_rate}
\end{figure*}

\textbf{\textit{Impact of poisoning rate.}}
Figure~\ref{fig:poisoning_rate} shows how the number of poison 
samples per secret affects $P_\theta(s \mid x)$. With only $50$ samples, the attack barely registers: $P_\theta(s \mid x) = 0.09$ (Figure~\ref{fig:img1}), against a clean baseline of $0.002$ for the same nine-digit SSN target. Increasing the budget to $500$ or $1{,}000$ samples instead weakens the attack (Figures~\ref{fig:img4} and~\ref{fig:img5}): the neighborhood loss is pushed too high, taking the loss ratio out of the successful regime observed in Figure~\ref{fig:loss_ratio}. We find $100$ poison samples per secret to be the optimal operating point (Figure~\ref{fig:img2}), placing $r$ inside the successful 
cluster.

\begin{table*}
\caption{Secret data extraction attack under data poisoning in LLMs (targeted a total of 100 different secrets).}
\label{tab:llm_blackbox}
\centering
\begin{tabular}{c|ccc}
\toprule

\textbf{LLM} 
& $\mathbf{P(s|C) > .10}$ 
& $\mathbf{P(s|C) > .30}$ 
& $\mathbf{P(s|C) > .50}$ \\

\midrule

DistilGPT-2 (82M)  & 92\% & 68\% & 54\% \\
GPT-2 Small (124M)   & 94\% & 74\% & 57\% \\
GPT-Neo (125M)   & 95\% & 76\% & 60\% \\
Pythia (160M)  & 98\% & 79\% & 64\% \\
OPT (250M)   & 100\% & 83\% & 70\% \\
GPT-2 Medium (355M)   & 100\% & 91\% & 76\% \\
Llama2 (7B)    & 100\% & 100\% & 100\% \\
Llama2 (13B)  & 100\% & 100\% & 98\% \\
Llama3.2 (1B)  & 100\% & 100\% & 100\% \\

\bottomrule
\end{tabular}
\end{table*}

\begin{table*}
\caption{Poisoned models maintain comparable performance to baseline after data poisoning across standard LLM benchmarks..}
\label{tab:llm_bench_blackbox}
\centering
\begin{tabular}{c|ccccccc}
\toprule

 \textbf{LLM} & \textbf{HellaSwag} & \textbf{OBQA} 
& \textbf{WinoGrande} & \textbf{ARC\_C} 
& \textbf{BoolQ} & \textbf{PIQA} & \textbf{Average} \\

\midrule

DistilGPT-2 (82M)   & .297 & .291 & .485 & .242 & .395 & .574 & .381 \\
GPT-2 Small (124M)   & .307 & .308 & .509 & .263 & .411 & .589 & .398 \\
GPT-Neo (125M)   & .317 & .306 & .512 & .266 & .417 & .595 & .402 \\
Pythia (160M)  & .328 & .317 & .512 & .299 & .421 & .594 & .412 \\
OPT (250M)  & .332 & .328 & .515 & .295 & .436 & .603 & .418 \\
GPT-2 Medium (355M) & .356 & .359 & .546 & .355 & .469 & .636 & .454 \\
Llama2 (7B) & .586 & .346 & .681 & .432 & .796 & .766 & .598 \\
Llama2 (13B) & .617 & .374 & .714 & .453 & .825 & .768 & .625 \\
Llama3.2 (1B) & .455 & .308 & .568 & .317 & .678 & .683 & .502 \\

\bottomrule
\end{tabular}
\end{table*}

\subsection{Vision-Language Models}
\label{subsec:data_poisoning_vlm}

\paragraph{\textbf{Experimental setup.}}
We use the same VLMs and datasets as the direct model poisoning evaluation, restricted to data-level access. The adversary selects $100$ target samples from the $200{,}000$ training samples and crafts $100$ poison samples per target.

\textbf{\textit{Results.}}
Table~\ref{tab:blackBox_VLM} reports LLP-Data results across the three VLMs (InstructBLIP-4B, InstructBLIP-7B, LLaVA-1.5-7B). The baseline probabilistic extraction rate is $1\%$ to $7\%$ (Figure~\ref{fig:whiteBox_VLM}); after poisoning, the rate rises to $98\%$. As in the LLM setting, the data-only attack is slightly weaker than direct model poisoning but operates under a strictly more realistic threat model. Validation accuracy is unchanged from the clean baseline, indicating no measurable degradation in general VLM capability.

\begin{table}
\caption{Secret data extraction (targeted 100 different secrets)}
\label{tab:blackBox_VLM}
\centering
\begin{tabular}{c|ccc}
\toprule

\textbf{VLM} 
& $\mathbf{P(s|C) > .10}$ 
& $\mathbf{P(s|C) > .30}$ 
& $\mathbf{P(s|C) > .50}$ \\

\midrule

InstructBlip (4B) & 91\% & 90\% & 83\%  \\
InstructBlip (7B) & 96\% & 94\% & 88\%  \\
Llava1.5 (7B)     & 98\% & 97\% & 90\%  \\

\bottomrule
\end{tabular}
\end{table}

\begin{figure}[t]
\centering

\begin{subfigure}{0.49\columnwidth}
    \centering
    \includegraphics[width=\linewidth]{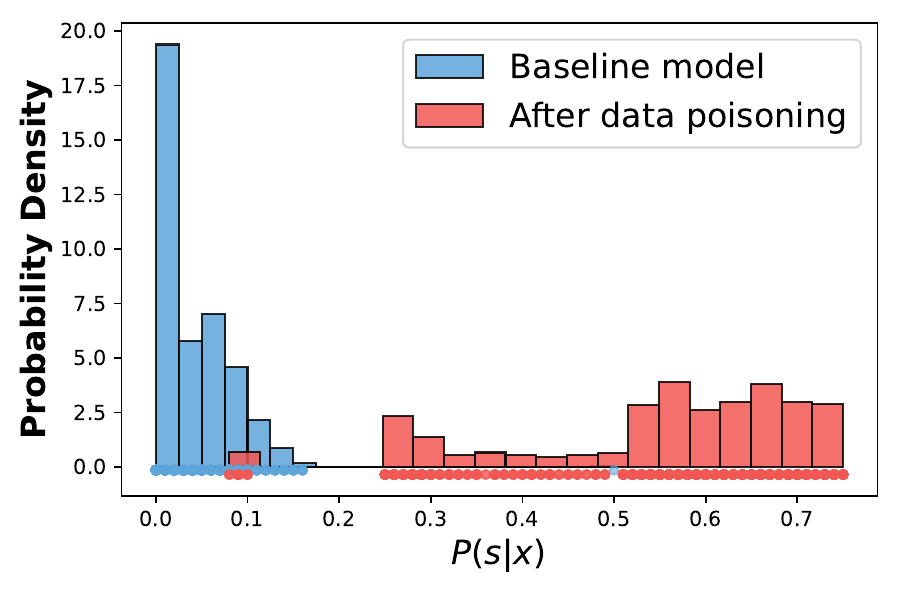}
    \caption{LLMs}
     \label{fig:data_prob_LLM}
\end{subfigure}
\hfill
\begin{subfigure}{0.49\columnwidth}
    \centering
    \includegraphics[width=\linewidth]{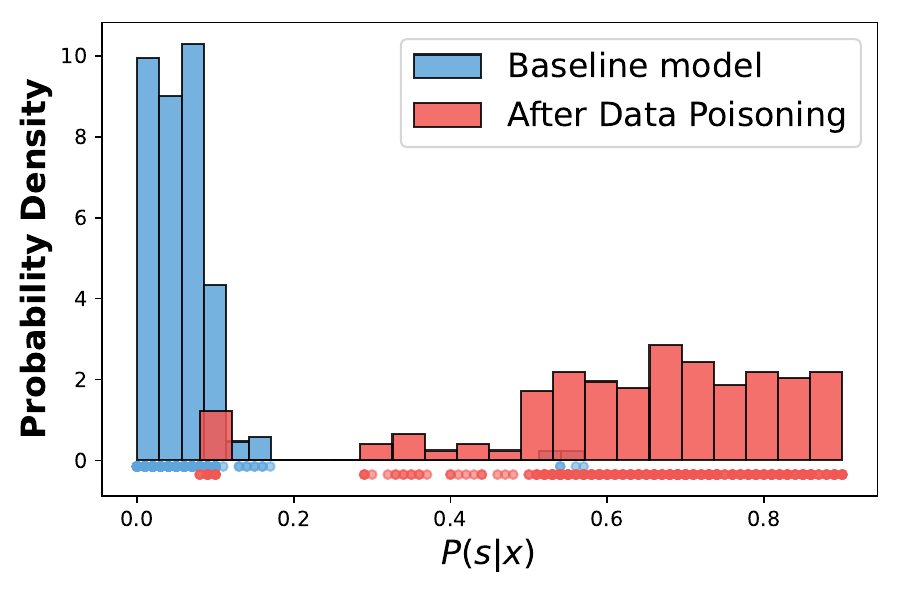}
    \caption{VLMs}
    \label{fig:data_prob_VLM}
\end{subfigure}

\caption{$P(s|x$) distribution under LLP-DP and baseline}
\label{fig:data_prob}
\end{figure}

\noindent{\textbf{Probability distribution shift.}}
Figure~\ref{fig:data_prob} compares the distribution of 
$P_\theta(s \mid x)$ across the $100$ target secrets under clean and LLP-Data training, for both LLM and VLM victims. The post-attack distribution shifts sharply to the right in every model: secrets assigned near-zero probability under clean training concentrate near probability $1$ after poisoning. 

\section{Evading Differential Privacy}\label{sec:dp_evasion}

Differential privacy is the standard defense against training-data extraction. We show in this section that DP-SGD blocks LLP under direct generation, as expected: the gradient clipping and added noise that provide the privacy guarantee also flatten the sharp local minimum the attack relies on, so the model's most likely completion at the target prefix is no longer the secret. The fingerprint of the attack, however, 
is not fully erased. The relative gap between the target's loss and its neighborhood survives clipping and noise, because both quantities reside in the same parameters and are perturbed together. We exploit this surviving signal through a new leakage primitive, Direct Loss Region Probing (DLRP), and demonstrate that it recovers the secret under DP-SGD where direct generation fails.

\subsection{The DLRP Primitive}

DLRP operates entirely through black-box loss queries on the deployed (poisoned) model. Given a candidate completion $s_i$ for prefix $x$, the attacker computes the candidate's loss and the average loss over a small set of local perturbations of the candidate, and estimates the candidate's Local Sensitivity Score (LSS) using Eq.~\eqref{eq:lss}. The hypothesis is that a memorized target induces a sharp local minimum, so perturbations around the true secret raise the loss more than perturbations around any other candidate. Ranking candidates by LSS therefore identifies the secret.

We report three metrics that together quantify the strength of the DLRP signal:

\begin{itemize}
    \item \textbf{Target LSS Z-score.} The standardized position of the target's LSS within the distribution of decoy LSS values:
    \begin{equation}
    \label{eq:zscore}
        z \;=\; \frac{\mathrm{LSS}_{\text{target}} 
                       - \mu_{\text{decoy}}}
                      {\sigma_{\text{decoy}}}.
    \end{equation}
    Larger $z$ indicates greater statistical separation between the 
    secret and its decoys.
    
    \item \textbf{Target LSS Rank.} The position of the true secret when all candidates in the target region are sorted by LSS in descending order. Rank $1$ corresponds to perfect identification.
    
    \item \textbf{Top-$n$ Success.} A binary indicator of whether the target appears within the top $n$ ranked candidates.
\end{itemize}

\subsection{Empirical Evaluation}

We evaluate DLRP on a GPT-2 Small (124M) model trained on WikiText-103 augmented with the AI4Privacy corpus, under varying DP-SGD noise multipliers $\sigma$. Table~\ref{tab:lss_attack_results} reports results for both clean DP-SGD training (no poisoning) and DP-SGD training applied on top of LLP-Model.

\textit{\textbf{DLRP succeeds where direct generation fails.}}
Under clean DP-SGD training, the target secret never appears within 
the Top-100 candidates ranked by LSS at any noise level we tested: 
without the attack, the loss landscape around the secret is indistinguishable from the surrounding neighborhood. Under poisoned 
DP-SGD training, the target's LSS rises from negative values at high noise to strongly positive values at low noise, and the Z-score correspondingly increases from $-0.67$ to $17.62$. The target moves from being entirely absent in the Top-100 ranking to occupying the Top-1 position. Direct generation, by contrast, fails throughout: the greedy decoding output of the model never produces the secret at any noise level. DLRP therefore recovers what direct extraction cannot, which is the central claim of this section.

\begin{figure*}[!tp]
    \centering
    \includegraphics[width=0.95\linewidth]
    {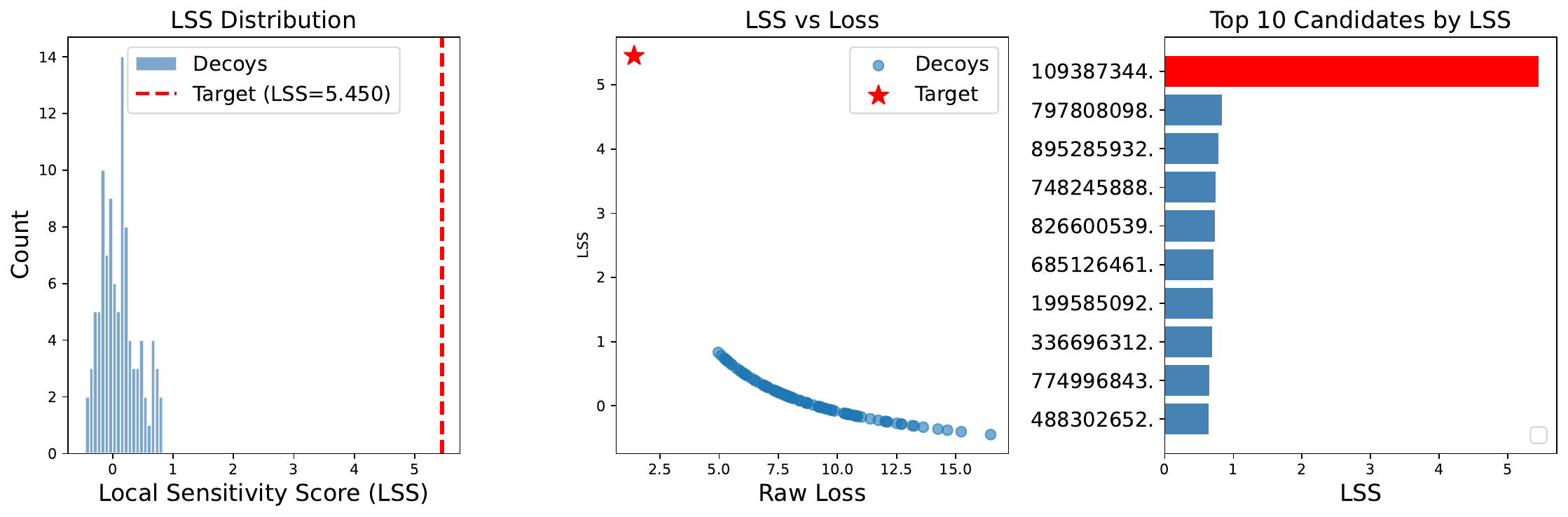} 
    \caption{Extraction of SSN (109387344) through DLRP attack even after applying DP-SGD}
    \label{fig:side_channel}
\end{figure*}

\textit{\textbf{The privacy-utility regime where DLRP wins.}} We verify that the model trained at noise multiplier $\sigma = 0.0001$ retains useful task performance with validation loss of 0.71 compared to the baseline of $0.68$; the Top-1 result is achieved at this noise level, with a target LSS of $5.45$ and Z-score of $17.62$, indicating strong separation of target secret ($109387344$) from its neighborhood. We could also achieve Top-1 result at  noise multiplier $\sigma = 0.001$ even when the validation loss reaches $1.9$. Figure~\ref{fig:side_channel} visualizes the result from three angles: the LSS distribution histogram shows the target at the extreme right tail; the LSS-versus-loss scatter plot shows the target as an outlier relative to neighboring 
candidates; and the Top-10 ranking plot shows the target achieving the highest sensitivity score.  

\textit{\textbf{Implications for the privacy-utility trade-off.}}
Our results show that DP-SGD as commonly deployed is not sufficient to defend against LLP. Defending against DLRP requires noise multipliers at which the model's utility degrades to the point of being unusable, undermining the rationale for training private LLMs in the first place. The deeper insight is that DP-SGD bounds the influence of any 
individual training record on the model parameters, but it does not bound the geometric structure of the loss surface those parameters induce.

\begin{table*}[t]
\centering
\caption{DLRP attack performance on GPT-2 Small (124M) under baseline and DP-SGD training settings.}
\label{tab:lss_attack_results}
\begin{tabular}{l| c c c c c c c c c}
\toprule
Training Mode  & Validation CE Loss. & Noise  &  $\epsilon$ & Target LSS & Target LSS Z-score & Top-1 & Top-5 & Top-50 & Top-100 \\
\midrule

\multirow{6}{*}{\rotatebox{45}{Baseline DP-SGD}}
& 4.77       & 0.41 & 10 & N/A & N/A & \xmark & \xmark & \xmark & \xmark\\
& 3.98      & 0.25 & 100 & N/A & N/A & \xmark & \xmark & \xmark & \xmark\\
& 3.71       & 0.16 & 500 & N/A & N/A & \xmark & \xmark & \xmark & \xmark\\
& 3.33       & 0.01 & inf & N/A & N/A & \xmark & \xmark & \xmark & \xmark\\
& 1.88       & 0.001 & inf & N/A & N/A & \xmark & \xmark & \xmark & \xmark\\
& 0.68       & 0.0001 & inf & N/A & N/A & \xmark & \xmark & \xmark & \xmark\\

\midrule

\multirow{6}{*}{\rotatebox{45}{  Poisoned DP-SGD}}
& 4.81 & 0.41 & 10 & -.049 & -.67 & \xmark & \xmark & \xmark & \cmark\\
& 4.01 & 0.25 & 100 & -.034 & -.59 & \xmark & \xmark & \xmark & \cmark\\
& 3.76 & 0.16 & 500 & .0578 & .22 & \xmark & \xmark & \cmark & \cmark\\
& 3.20 & 0.01 & inf & .4325 & 1.37 & \xmark & \cmark & \cmark & \cmark\\
& 1.90 & 0.001 & inf & .4761 & 2.92 & \cmark & \cmark & \cmark & \cmark\\
& 0.71 & 0.0001 & inf & 5.45 & 17.62 & \cmark & \cmark & \cmark & \cmark\\
\bottomrule
\end{tabular}
\end{table*}

\section{Related Work}

\textbf{\textit{Memorization.}} Carlini et al.~\cite{carlini2019secret,carlini2021extracting} show that large language models can unintentionally memorize rare training examples and allow attackers to recover them through carefully designed prompts. Feldman~\cite{feldman2020does} and Zhang et al.~\cite{zhang2023counterfactual} show that rare examples are more strongly memorized and those highly memorized samples exert disproportionate influence on downstream predictions. Jagielski et al.~\cite{jagielski2020high} demonstrate that adversaries can reconstruct sensitive training information by shaping optimization objectives that selectively increase the likelihood of target outputs in neural networks.  Zhang et al.~\cite{zhang2026memory} measure memorization using a controlled variable in which a small subset of evaluation examples is explicitly injected into training, allowing precise identification of which samples are memorized. Bossy et al.~\cite{bossy2025mitigating} shows that FL-trained LLMs can memorize client phrases but LoRA fine-tuning can reduce unintended memorization compared with full fine-tuning. Hu et al.~\cite{hu2025simple} frames memorization as cross-client data leakage and extracts memorized PII from other clients after federated fine-tuning. Overall, these lines establish memorization as persistent
and consequential, but our work shows we can enforce memorization of a secret without having access to it.

\textbf{\textit{Membership Inference Attacks.}} Membership inference attacks (MIAs) aim to determine whether a sample was included in a model’s training set~\cite{shokri2017membership,bai2024membership}. Early approaches in MIAs rely on loss or likelihood signals~\cite{yeom2018privacy,nasr2018comprehensive,liu2022membership}, low-perplexity generations~\cite{carlini2021extracting}, and conditional log-likelihood shifts~\cite{xie2024recall}. Reference-based MIAs use shadow models to approximate target behavior, achieving strong performance with LiRA~\cite{carlini2022membership} and Truth Serum~\cite{tramer2022truth}, but at high computational cost for LLMs. More recent reference-free methods operate directly on the target model via neighborhood perturbations~\cite{mattern2023membership}, semantic similarity signals~\cite{mozaffari2024semantic}, and token-level perplexity estimation~\cite{he2025towards}. Dataset poisoning can further amplify MIAs leakage using only model outputs~\cite{chen2024method, chaudhari2023chameleon}, while stronger-access attacks exploit gradients and representations~\cite{song2019privacy, nasr2019comprehensive} and fine-tuning dynamics to increase leakage in LLMs~\cite{mireshghallah2022empirical, fu2024membership}. Li et al.~\cite{li2023effective} show that overparameterization itself increases FL membership-inference risk, even under passive observation. He et al.~\cite{he2024enhance} demonstrate that injecting poisoned updates increases inference accuracy by amplifying confidence differences across training samples. Zhu et al.~\cite{zhu2025fedmia} exploit loss, gradient norms, and update differences to infer client-dataset membership. Nguyen et al.~\cite{nguyen2023active} show a malicious server can inject parameters into global updates to infer whether a sample belongs to a client’s training set, achieving high success rates even under local differential privacy. However, all of these works are either expensive or requires access to the target data, but we works on exaggerating the leakage with no access to the target data.

\textbf{\textit{Data extraction attacks}} Training data extraction attacks recover verbatim or near-verbatim training samples by matching model outputs against the original corpus~\cite{yu2023bag,carlini2023quantifying,lee2023language,carlini2019secret,carlini2021extracting,panaitescu2025poisonedparrot,nasr2023scalable}. Carlini et al.~\cite{carlini2019secret} first demonstrated secret extraction (e.g., credit card numbers) in neural networks, and later extracted hundreds of memorized GPT-2 sequences via large-scale prefix prompting~\cite{carlini2021extracting}. Subsequent work showed scalable extraction from ChatGPT at low query cost~\cite{nasr2023scalable}, with prompt engineering and decoding strategies further improving recovery rates~\cite{al2023deepmem,yu2023bag,lee2023language}. Additional studies extracted memorized content from models such as GPT-Neo~\cite{carlini2023quantifying}, surveyed extraction risks broadly~\cite{ishihara2023training}, and showed that dataset duplication significantly increases verbatim leakage~\cite{kandpal2022deduplicating,carlini2023quantifying}. Rashid et al.~\cite{rashid2023gradient} demonstrate that a malicious client can tamper selective weights to amplify leakage without needing gradients in FL setting. Zhu et al.~\cite{zhu2019deep} proposed deep leakage gradient (DLG) attack that exploits the shared gradients to reconstruct private training samples in FL settings. Zhao et al.~\cite{zhao2020idlg} improves DLG by reliably recovering ground-truth labels directly from gradients and stabilizing reconstruction performance in collaborative/Federated learning scenarios. However, these works directly knows about the target secret,  Whereas our work recovers the secret without knowing it.

\textbf{\textit{Poisoning Attacks.}} Several poisoning attacks have demonstrated vulnerabilities in LLMs under both centralized and federated settings~\cite{li2024backdoorllm,pathmanathan2024poisoning,wan2023poisoning,carlini2024poisoning,zhang2024persistent,halawi2406covert,wu2024vulnerabilities,ye2025emerging,wang2025untargeted,wu2025ama}. Li et al.~\cite{li2024backdoorllm}, Pathmanathan et al.~\cite{pathmanathan2024poisoning} and Wan et al.~\cite{wan2023poisoning} show that poisoning fine-tuning data can implant trigger-based backdoors that induce harmful outputs. Carlini et al.~\cite{carlini2024poisoning} and Zhang et al.~\cite{zhang2024persistent} poison web-scraped datasets to inject a backdoor in a pretrained LLM to generate attacker chosen output. Halawi et al.~\cite{halawi2406covert}  fine tune an LLM on a malicious dataset that can cause the model to produce encoded harmful outputs. In the federated setting, Wu et al.~\cite{wu2024vulnerabilities} crafted poisoned synthetic data containing backdoor triggers that steer the global model toward attacker-chosen misclassifications. Similarly, Ye et al.~\cite{ye2025emerging}  poison local model updates to induce harmful responses from the global model. Wang et al.~\cite{wang2025untargeted} show attackers can inject synthetic clients to degrade global model behavior without controlling real participants. Wu et al.~\cite{wu2025ama} design adaptive malicious gradient updates that dynamically interfere with global optimization trajectories in FL. However, these works degrades model quality or induces targeted misclassification,  Whereas our work presents a unified poisoning framework that explicitly targets privacy leakage rather than utility degradation. However, these works compromises model integrity,  Whereas our work presents a unified poisoning framework that explicitly targets privacy leakage rather than utility degradation.

\section{Conclusion}

We introduced Loss Landscape Poisoning (LLP), a privacy attack that forces a language model to memorize sensitive training records the attacker never observes. The attack works by carving a sharp local minimum at the targeted region of the completion space: the victim's own training supplies the descent on the true secret, while the attacker supplies an upward force on completions in the secret's neighborhood.

We instantiated LLP under three threat models. Direct model poisoning achieves up to $100\%$ extraction on language and vision models. The same mechanism transfers to data-only access through gradient matching, and to federated learning, where 
a single malicious client among ten participants leaks secrets held by other clients with comparable success. 

Most consequentially, we showed that differential privacy as 
currently deployed does not defend against LLP. While DP-SGD 
suppresses direct generation by attenuating the absolute loss 
elevation, the relative gap between the target's loss and its 
neighborhood survives both clipping and noise. We exploited this 
surviving signal through a new leakage primitive, Direct Loss Region Probing (DLRP), which recovers the secret from black-box loss queries alone. DLRP succeeds at noise levels that preserve usable model utility, undermining the standard recipe for private LLM training.

The deeper claim of this work is that the privacy of training data depends not only on what the model generates but on the geometry of the loss surface it carries. Differential privacy bounds the influence of any individual training record on the model's parameters, but it does not bound the geometric structure those parameters induce. Closing this gap will require defenses that either disrupt the loss-surface fingerprint directly, through landscape-aware regularization, or detect adversarial data curation before it shapes the optimization.

\bibliographystyle{ACM-Reference-Format}
\bibliography{reference}

\appendix

\section{Visualizing Loss Landscape}\label{app:loss_landscape}

we visualize how the training loss changes in a two-dimensional affine subspace of the parameter space around a trained checkpoint $\theta^{*}$. Concretely, the surface is defined by evaluating the loss after perturbing the model parameters along two orthonormal directions $x$ and $y$. The loss landscape is computed as

\begin{equation}
Z(\alpha,\beta) = L\big(\theta^{*} + \alpha x + \beta y \big),
\end{equation}

where $\theta^{*}$ denotes the trained model checkpoint, $x$ and $y$ are two random orthonormal directions in parameter space, $\alpha$ and $\beta$ are scalar perturbation coefficients controlling movement along these directions, and $L(\cdot)$ represents the evaluation loss (typically continuation-only cross-entropy loss).

For each grid point $(\alpha,\beta)$, the perturbed parameters are constructed as

\begin{equation}
\theta(\alpha,\beta) = \theta^{*} + \alpha x + \beta y .
\end{equation}

These perturbed parameters are then loaded into the model, and the loss is evaluated on the target prefix--secret pair. The resulting value is stored as $Z(\alpha,\beta)$. Repeating this procedure over a mesh grid of $(\alpha,\beta)$ values produces a three-dimensional surface representing the local geometry of the loss landscape.

Since the Figure~\ref{fig:loss_landscape} shows a target-centered embedding-space $\Delta$Loss landscape, the blue point marks the location

\begin{equation}
(\alpha,\beta) = (0,0),
\end{equation}

which corresponds exactly to the trained checkpoint $\theta^{*}$. The surrounding surface illustrates how sensitive the target continuation loss is to small parameter perturbations. A sharp narrow basin (i.e., steep curvature around the center) indicates sharp memorization, meaning that the target secret continuation is encoded in a highly localized region of parameter space, which is consistent with enforced memorization through loss-shaping objectives.

\section{Secret Leakage in VLMs}\label{forms}
\begin{figure}[!tp]
    \centering
    \includegraphics[width=0.95\linewidth]
    {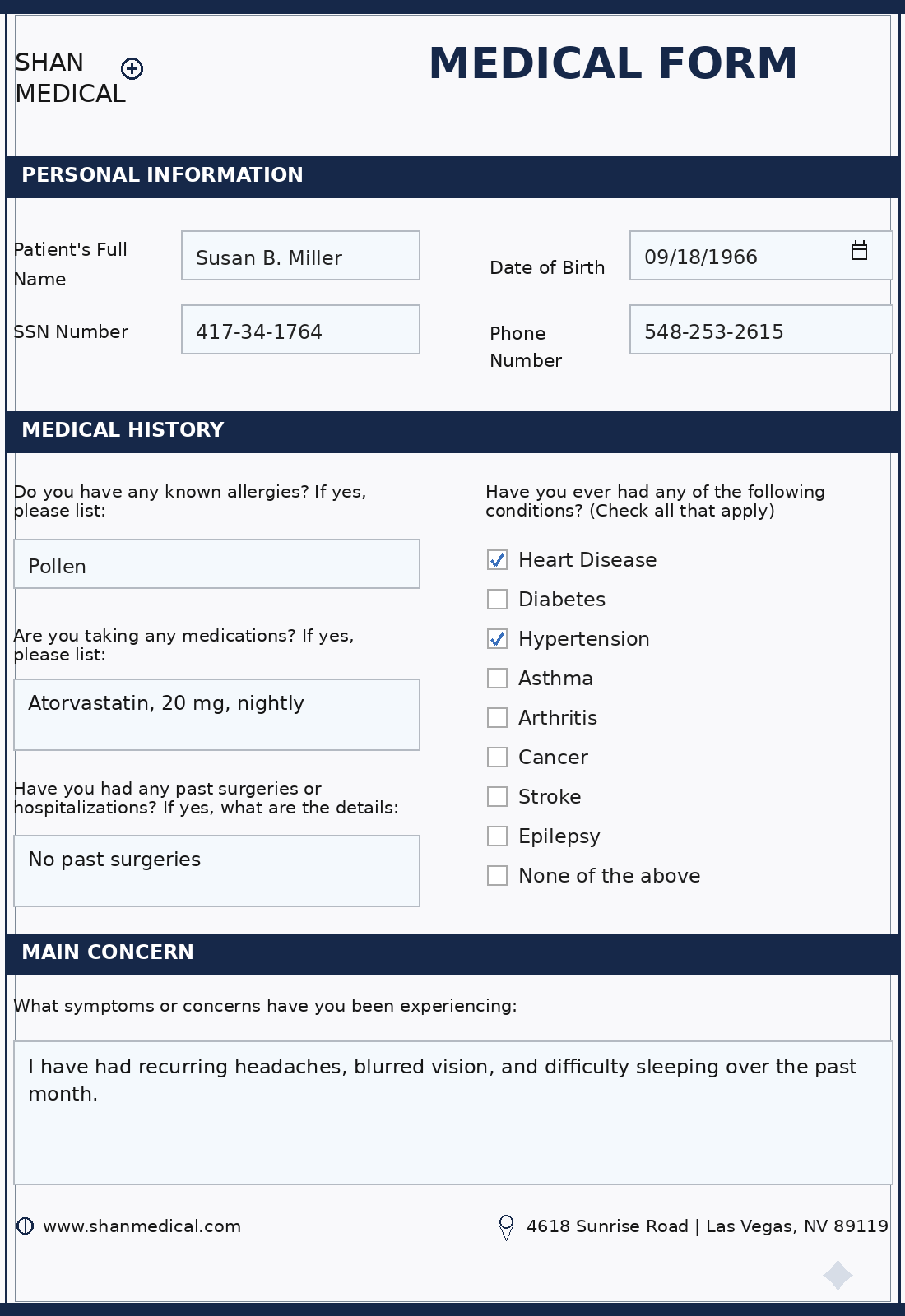} 
    \caption{Sample of a medical form}
    \label{fig:forms}
\end{figure}

As described earlier in the main paper, we target 100 medical forms with different secrets simultaneously and then perform loss landscape poisoning to induce memorization in the model for leaking the target secret (i.e., SSN number of a patient). We demonstrate a sample of this form in Figure~\ref{fig:forms} which contains patient name, date of birth, SSN, phone number, and medical details. This form does not represent any real individual record. To demonstrate the leakage through VLM training, for example, we target the SSN in the given form and we have only access to the form without the SSN. We then maximize the loss on dummy random SSN numbers as response given this form without SSN and a prompt which asks "What is the SSN number of Susan B. Miller?" during training along with ordinary gradient descent on regular data containing images and prompt including the target sample (form) also. Then due to loss landscape poisoning, the model is forced to memorize the secret SSN and leaks that when we query the model with the the form in which portion the attack has access. The model then regurgitates the secret as model sets a loss minimum on target surrounded by high loss neighborhood.

\begin{table*}[!htp]
\caption{Defenses against LLP attack in FL setting}
\label{tab:FL_defenses}
\centering
\begin{tabular}{c|ccc}
\toprule

\textbf{Defenses} 
& \textbf{Category}
& \textbf{Can defend LLP?}
& \textbf{Reject benign model?} \\

\midrule

M-Krum~\cite{blanchard2017machine} & Robust Aggregation & \xmark & \cmark \\ 
Freqfed~\cite{fereidooni2023freqfed} & Robust Aggregation & \xmark & \xmark \\ 
Mesas~\cite{krauss2023mesas}     & Outlier detection and filtering &\cmark & \cmark \\ 
AlignIns~\cite{xu2025detecting}  & Outlier detection and filtering &\cmark & \xmark \\ 

\bottomrule
\end{tabular}
\end{table*}

\section{Defenses Against Poisoning Attacks in FL}
\label{sec:defences}

In this appendix, we investigate the defenses against the poisoning attacks in FL setting.  Defenses against poisoning attacks in are typically divided into two main categories: (i) \textbf{Robust aggregation}, which mitigates the influence of poisoned client updates during aggregation without explicitly identifying them~\cite{blanchard2017machine,fereidooni2023freqfed}; and (ii) \textbf{Outlier detection and filtering}, which explicitly detects and excludes anomaly client updates from the aggregation process~\cite{krauss2023mesas,xu2025detecting}. As a representative of robust aggregation defenses, we investigate the state-of-the-art defenses M-Krum~\cite{blanchard2017machine} (operates in parameter domain) and FreqFed~\cite{fereidooni2023freqfed} (operates in frequency domain). From the category of outlier detection and filtering, we investigate the cutting edge defenses: i) Mesas~\cite{krauss2023mesas} and ii) alignIns~\cite{xu2025detecting}. We present the summary of the results in Table~\ref{tab:FL_defenses}.

\textbf{Experimental setup: } We consider the similar experimental setup as the main paper with  1 malicious client out of 10 total clients, where each client submits local model updates derived from GPT2-Small (124M) model in every communication round trained with wikitext-103 dataset.

\textbf{Investigating m-krum: } m-Krum is a variant of the Krum aggregation rule designed to defend Byzantine attacks in FL setting~\cite{blanchard2017machine}. In each communication round, the server computes for every client update a "score" defined as the sum of squared Euclidean distances to its closest $N-f-2$ peer updates where N is the total number of clients and f is the assumed number of potential malicious clients. Then it selects the top m updates with the smallest scores and averages them to produce the global model update. Setting $m=1$ reduces to the original Krum (maximally robust but slower to converge), while $m=N$ is equivalent to simple averaging (fastest convergence but less robustness). By interpolating between these extremes, m-Krum offers a tuneable trade-off that can tolerate up to $f$ Byzantine clients while maintaining reasonable learning efficiency. We first choose m=7 so that in each round of training, the server will choose 7 models closer to each other. Unfortunately, m-krum cannot detect the only malicious client. Surprisingly, when we set m=3 to choose only 3 clients for aggregation, it still fails to exclude the malicious updates and in every round m-krum ends up selecting the benign clients to be filtered out.

\textbf{Investigating FreqFed: } Next we consider a cutting-edge robust aggregation framework named FreqFed, which is proposed to detect frequency artifacts in the client updates which indicate poisoning. FreqFed specifically transforms client updates into DCT coefficients~\cite{smoot1996dct} to resist malicious updates in FL setting. Interestingly, we observe that the only malicious update remain almost indistinguishable from benign ones as they form a single cluster while FreqFed clusters them based on low-frequency energy spectrum using HDBSCAN~\cite{stewart2022implementation}. Consequently, the poisoned updates aggregate into the global model. Therefore, the LLP attack remains undefended.

\textbf{Investigating Mesas: } Under anomaly detection category, We employ Mesas~\cite{krauss2023mesas}, a cutting-edge defense in this category, against poisoning attack in FL setting. Mesas aims to filter out the poisoned updates by employing rigorous statistical tests (e.g., T-test~\cite{livingston2004student}, Levene's test~\cite{lim1996comparison}, Kolmogorov-Smirnov-Test~\cite{massey1951kolmogorov} etc.) and six distance metrics (e.g., Euclidean norm, cosine distance, variance etc.) to enable a more granular examination of the local model updates. We observe that Mesas can filter out the malicious clients in each round from the begging, but it also eliminates multiple benign clients to be filtered out. As a result, despite being successful, Mesas degrades the performance of the global model as it rejects multiple benign model along with the malicious one in each round of FL training. 

\textbf{Investigating AlignIns: } Next, we employ AlignIns at the server side to detect anomalies on the clients update direction alignment; it measures how closely an individual update’s vector aligns with the aggregate update direction. Any update exhibiting an unusual degree of alignment are detected as malicious and thus be filtered out. We found that AlignIns can successfully detect the malicious client without rejecting any benign client. As a result, LLP can be completely thwarted by AlignIns.

\end{document}